\documentclass[a4paper]{article}
\usepackage{a4wide}

\usepackage{lmodern}
\usepackage{amsmath, amssymb, amstext}
\usepackage{booktabs}
\usepackage{authblk}
\usepackage{graphicx}
\usepackage{subfig}
\usepackage{hyperref}
\usepackage[numbers]{natbib}
\graphicspath{%
  {./metapost/}
  {./images_REV/}
  {./images_REV2/}
}

\newcommand{\R}{\mathbb{R}}
\newcommand{\abs}[1]{\left| {#1} \right|}

\renewcommand{\vec}[1]{\mathbf{#1}}
\newcommand{\tensor}[1]{\mathbf{#1}}
\newcommand{\isotensor}[1]{\overline{\mathbf{#1}}}
\newcommand{\vu}[0]{\vec{u}}
\newcommand{\vx}[0]{\vec{x}}
\newcommand{\vn}[0]{\vec{n}}
\newcommand{\ve}[0]{\vec{e}}

\newcommand{\intd}[1]{\;\mathrm{d}#1}

\newcommand{\dt}[0]{\intd{t}}
\newcommand{\dx}[0]{\intd{x}}
\newcommand{\dvx}[0]{\intd{\vx}}

\newcommand{\diff}[2]{\frac{\mathrm{d}#1}{\mathrm{d}#2}}

\newcommand{\subscript}[2]{{{#1}_{#2}}}
\newcommand{\txtsubscript}[2]{\subscript{#1}{\text{#2}}}

\newcommand{\txtfrac}[2]{\ensuremath{\frac{\text{#1}}{\text{#2}}}}

\newcommand{\sphh}[0]{\ensuremath{\mathrm{\mathbf{Sph}_{5}}}}
\newcommand{\sphH}[0]{\ensuremath{\mathrm{\mathbf{Sph}_{25}}}}
\newcommand{\sphHH}[0]{\ensuremath{\mathrm{\mathbf{Sph}_{150}}}}

\newcommand{\lva}[0]{\ensuremath{\mathrm{\mathbf{LV}_{A}}}}
\newcommand{\lvb}[0]{\ensuremath{\mathrm{\mathbf{LV}_{B}}}}
\newcommand{\lvc}[0]{\ensuremath{\mathrm{\mathbf{LV}_{C}}}}
\newcommand{\lvd}[0]{\ensuremath{\mathrm{\mathbf{LV}_{D}}}}

\newcommand{\lvagu}[0]{\ensuremath{\mathrm{\mathbf{LV}_{A-Gu}}}}
\newcommand{\lvbgu}[0]{\ensuremath{\mathrm{\mathbf{LV}_{B-Gu}}}}
\newcommand{\lvcgu}[0]{\ensuremath{\mathrm{\mathbf{LV}_{C-Gu}}}}
\newcommand{\lvdgu}[0]{\ensuremath{\mathrm{\mathbf{LV}_{D-Gu}}}}
\newcommand{\lvadem}[0]{\ensuremath{\mathrm{\mathbf{LV}_{A-Dem}}}}
\newcommand{\lvbdem}[0]{\ensuremath{\mathrm{\mathbf{LV}_{B-Dem}}}}
\newcommand{\lvcdem}[0]{\ensuremath{\mathrm{\mathbf{LV}_{C-Dem}}}}
\newcommand{\lvddem}[0]{\ensuremath{\mathrm{\mathbf{LV}_{D-Dem}}}}

\newcommand{\plv}{\ensuremath{p_{\mathrm{LV}}}}
\newcommand{\pao}{\ensuremath{p_{\mathrm{ao}}}}
\newcommand{\qpeak}{\ensuremath{\hat{q}_{\mathrm{LV}}}}
\newcommand{\plvpeak}{\ensuremath{\hat{p}_{\mathrm{LV}}}}
\newcommand{\paopeak}{\ensuremath{\hat{p}_{\mathrm{ao}}}}
\newcommand{\tplvpeak}{\ensuremath{t_{\hat{p}}}}
\newcommand{\tqlvpeak}{\ensuremath{t_{\hat{q}}}}
\newcommand{\tejs}{\ensuremath{t_{\rm{0,ej}}}}

\newcommand{\tES}{\ensuremath{t_{\mathrm{ES}}}}
\newcommand{\tED}{\ensuremath{t_{\mathrm{ED}}}}
\newcommand{\Tsys}{\ensuremath{T_{\mathrm{sys}}}}
\newcommand{\Tsct}{\ensuremath{T_{\text{sct}}}}
\newcommand{\tpowerpeak}{\ensuremath{t_{\rm{pow,peak}}}}

\newcommand{\strain}{\ensuremath{\boldsymbol{\varepsilon}}}
\newcommand{\straindot}{\ensuremath{\dot{\boldsymbol{\varepsilon}}}}
\newcommand{\sigmatot}[0]{\boldsymbol{\sigma}}
\newcommand{\sigmatotsph}[0]{\txtsubscript{\boldsymbol{\sigma}}{sph}}
\newcommand{\sigmapas}[0]{\txtsubscript{\boldsymbol{\sigma}}{pas}}
\newcommand{\sigmaact}[0]{\txtsubscript{\boldsymbol{\sigma}}{act}}

\newcommand{\sigmaLstar}[0]{\txtsubscript{\sigma}{L,\ensuremath{\star}}}
\newcommand{\sigmaLext}[0]{\txtsubscript{\sigma}{L,H}}
\newcommand{\sigmaLsmp}[0]{\txtsubscript{\sigma}{L,h}}
\newcommand{\sigmaLvol}[0]{\txtsubscript{\sigma}{L,V}}

\newcommand{\sigmacirc}[0]{\txtsubscript{\sigma}{circ}}
\newcommand{\sigmatht}[0]{\subscript{\sigma}{\theta\theta}}
\newcommand{\sigmaphi}[0]{\subscript{\sigma}{\varphi\varphi}}
\newcommand{\sigmarad}[0]{\subscript{\sigma}{rr}}

\newcommand{\msigmatht}[0]{\subscript{\overline{\sigma}}{\theta\theta}}
\newcommand{\msigmaphi}[0]{\subscript{\overline{\sigma}}{\varphi\varphi}}
\newcommand{\msigmarad}[0]{\subscript{\overline{\sigma}}{rr}}
\newcommand{\msigmacirc}[0]{\txtsubscript{\overline{\sigma}}{circ}}
\newcommand{\msigmamean}[0]{\txtsubscript{\overline{\sigma}}{mean}}

\newcommand{\powerint}[0]{\txtsubscript{P}{int}}
\newcommand{\powerintdensity}[0]{\txtsubscript{p}{int}}
\newcommand{\powerintpeak}[0]{\txtsubscript{\hat{P}}{int}}

\newcommand{\powerintstar}[0]{\txtsubscript{P}{int,\ensuremath{\star}}}

\newcommand{\powerintsmp}[0]{\txtsubscript{P}{int,h}}

\newcommand{\powerintext}[0]{\txtsubscript{P}{int,H}}

\newcommand{\powerintvol}[0]{\txtsubscript{P}{int,V}}

\newcommand{\powerext}[0]{\txtsubscript{P}{ext}}
\newcommand{\powerextpeak}[0]{\txtsubscript{\hat{P}}{ext}}

\newcommand{\powerextmeanao}[0]{\txtsubscript{\bar{P}}{ext,ao}}
\newcommand{\powereffclin}[0]{\txtsubscript{P}{eff,clin}}
\newcommand{\powereff}[0]{\txtsubscript{P}{eff}}
\newcommand{\ihp}{\ensuremath{\text{IHP}}}
\newcommand{\ehp}{\ensuremath{\text{EHP}}}
\newcommand{\ihw}{\ensuremath{\text{IHW}}}
\newcommand{\MAP}{\ensuremath{\text{MAP}}}
\newcommand{\CO}{\ensuremath{\text{CO}}}

\newcommand{\workint}[0]{\txtsubscript{W}{int}}
\newcommand{\workintdensity}[0]{\txtsubscript{w}{int}}
\newcommand{\workext}[0]{\txtsubscript{W}{ext}}

\newcommand{\workintstar}[0]{\txtsubscript{W}{int,\ensuremath{\star}}}
\newcommand{\workintsmp}[0]{\txtsubscript{W}{int,h}}
\newcommand{\workintext}[0]{\txtsubscript{W}{int,H}}
\newcommand{\workintvol}[0]{\txtsubscript{W}{int,V}}

\newcommand{\volcav}[0]{\txtsubscript{V}{cav}}
\newcommand{\volmyo}[0]{\txtsubscript{V}{myo}}

\renewcommand{\log}[0]{\operatorname{log}}
\renewcommand{\exp}[0]{\operatorname{exp}}
\newcommand{\tr}[0]{\operatorname{tr}}

\newcommand{\sectionend}[1]{}
\newcommand{\subsectionend}[1]{}
\newcommand{\subsubsectionend}[1]{}

\makeatletter
\def\blfootnote{\gdef\@thefnmark{}\@footnotetext}
\makeatother

\author[1]{Matthias A.F. Gsell}
\author[1,4]{Christoph M. Augustin}
\author[1]{Anton J. Prassl}
\author[1]{Elias Karabelas}
\author[3]{Joao F. Fernandes}
\author[2,3]{Marcus Kelm}
\author[3]{Leonid Goubergrits}
\author[2,3]{Titus Kuehne}
\author[1, $\dagger$]{Gernot Plank}

\affil[1]{Institute of Biophysics, Medical University of Graz, Graz, Austria}
\affil[4]{Shadden Research Group, Department of Mechanical Engineering, University of California, Berkeley, CA, USA}
\affil[2]{Department of Congenital Heart Disease/Pediatric Cardiology, German Heart Institute Berlin, Berlin, Germany}
\affil[3]{Institute for Imaging Science and Computational Modelling in Cardiovascular Medicine, Charit\'{e}-Universit\"atsmedizin Berlin, Berlin, Germany}
\affil[$\dagger$]{Correspondance: Gernot Plank, Medical University of Graz,
  Institute of Biophysics, Neue Stiftingtalstrasse 6/IV, 8010 Graz, Austria. gernot.plank@medunigraz.at}
\date{}
\title{Assessment of wall stresses and mechanical heart power in the left ventricle:
Finite element modeling versus Laplace analysis
\protect\thanks{This research was supported by the grants F3210-N18 and I2760-B30
from the Austrian Science Fund (FWF) and the EU grant CardioProof 611232
and a Marie Sk{\l}odowska--Curie fellowship (GA 750835) to CA.}}

\begin{document}

\maketitle

\begin{abstract}
  \emph{Introduction:} Stenotic aortic valve disease (AS) causes pressure overload of the left ventricle (LV)
  that may trigger adverse remodeling and precipitate progression towards heart failure (HF).
  As myocardial energetics can be impaired during AS,
  LV wall stresses and biomechanical power provide a complementary view of LV performance
  that may aide in better assessing the state of disease.
  \emph{Objectives:} Using a high-resolution electro-mechanical (EM) \emph{in silico} model of the LV
  as a reference,
  we evaluated clinically feasible Laplace-based methods for assessing global LV wall stresses and biomechanical power.
  \emph{Methods:} We used $N=4$ \emph{in silico} finite element (FE) EM models of LV and aorta
  of patients suffering from AS.
  All models were personalized with clinical data under pre-treatment conditions.
  LV wall stresses and biomechanical power were computed accurately from FE kinematic data
  and compared to Laplace-based estimation methods
  which were applied to the same FE model data.
  \emph{Results and Conclusion:}
  Laplace estimates of LV wall stress are able to provide a rough approximation
  of global mean stress in the circumferential-longitudinal plane of the LV.
  However, according to FE results
  spatial heterogeneity of stresses in the LV wall is significant,
  leading to major discrepancies between local stresses and global mean stress.
  Assessment of mechanical power with Laplace methods is feasible,
  but these are inferior in accuracy compared to FE models.
  The accurate assessment of stress and power density distribution in the LV wall
  is only feasible based on patient-specific FE modeling.

\end{abstract}

\blfootnote{\textbf{Abbreviations:} AVD, aortic valve disease; LV, left ventricle;
HF, heart failure; PV, pressure-volume; EM, electro-mechanical; IHP, internal mechanical heart power;
FE, finite element; IVC, isovolumetric contraction; IVR, isovolumetric relaxation.}


\section{Introduction}
  \label{sec:introduction}
  In AS elevated pressure gradients impose a higher load upon the LV.
  Under such conditions, the pressure produced by the LV
  must increase in order to achieve an adequate cardiac output that meets the metabolic demands.
  This requires the LV wall to generate higher active forces,
  which can be achieved either by an increase in wall stresses
  or a change in ventricular shape and mass.
  Such pressure overload conditions, if persistent
  for long enough, trigger adverse remodeling processes,
  eventually precipitating progression towards HF \cite{grossman13:_myocardial}.
  Treatments aim at alleviating pressure overload by reducing transvalvular pressure gradients
  closer to normal levels by surgical or catheter based aortic valve replacement \cite{ali18:_tavi}.
  However, re-stenosis frequently occurs and despite a successful reduction of transvalvular pressure gradients,
  a majority of patients remains hypertensive, consequently showing increased risk for irreversible course of HF
  and higher morbidity and mortality \cite{elhmidi14:_low_gradient}.
  Thus, a successful reduction of pathologically elevated pressure gradients alone
  cannot be considered a reliable prognostic marker of long-term post-treatment outcomes
  in these patient cohorts.


  As a consequence, alternative biomarkers beyond pressure gradients are sought to
  that provide a complementary view of cardiac function
  and, potentially, offer a higher predictive power with regard to
  outcomes.
  In a recent study, the use of end-diastolic or end-systolic wall stresses
  as assessed by a wall stress index has been proposed as a novel diagnostic criterion
  of HF \cite{alter16:_wallstress}.
  This is physiologically motivated as elevated wall stress levels
  are assumed to impair the balance between metabolic supply and demand
  \cite{strauer79:_oxygen_wall_stress} by hindering perfusion and, thus, contribute
  towards adverse remodeling \cite{aikawa01:_wall_stress_remodelling}.
  Wall stresses are directly linked to the mechanical power
  generated by the myocardial muscle
  and the work performed by it and as such can be considered a metabolic marker.
  Different approaches have been proposed to assess work and the energy expenditure
  of the myocardium. As a direct measurement of energy metabolism Positron Emission
  Tomography (PET) was used \cite{gucclu2015myocardial, hansson2017myocardial},
  however the method is limited due to its complexity including the need for tracers
  involving ionizing radiation. More recently, a concept of biomechanical internal
  myocardial heart power ($\ihp$), necessary to maintain adequate cardiac output
  ($\ehp$, external heart power), has been introduced in patients with aortic coarctation
  \cite{fernandes17:_beyond}. Findings in this cohort suggest that the ratio $\ehp/\ihp$,
  referred to as power efficiency, improved mostly in those cases with elevated $\ihp$.
  While potential marker qualities of such concepts need to be further evaluated, it
  has remained a yearned-for goal to lower energy expenditure and increase efficiency
  of the myocardium likewise in any treatment procedure, including those for stenotic
  valvular and vascular disease.

  Another method used to determine the work performed by the muscle is pressure-volume (PV) relations.
  These are usually measured using conductance catheter techniques.
  However, these procedures are invasive, time-consuming and expensive.
  Alternatively, PV-loops are measured with 3D echo or MRI \cite{schmitt2016cardiac},
  but even these methods are complex and
  pressure and volume traces are not recorded simultaneously.

  Despite the diagnostic potential of markers based on wall stresses and
  expended mechanical power,
  this assessment has not evolved towards a routinely used diagnostic tool in the
  clinic mainly due to methodological limitations.
  Attempts to address this
  relied upon different variants of Laplace's law, which require the
  acquisition of only a small number of measures
  representing LV cavity volume, wall width and pressure
  \cite{alter16:_wallstress,fernandes17:_beyond}.
  However, these approaches are based on simplifications with regard to
  LV geometry, tissue structure and biomechanical properties
  as they assume the LV a thin-walled mechanically isotropic spherical shell.
  The accuracy and validity of these simplifications have not been firmly established,
  thus casting doubt on the reliability and fidelity of any metrics based on them \cite{moriarty1980law,zhang2011comparison}.
  Experimental validation based on direct measurements of stresses \emph{in vivo}
  is challenging and not feasible yet with currently available technologies.
  However, an indirect inference is viable using computational tools
  such as FE modeling where 3D wall stresses can be computed
  from a set of reliable strains
  -- either measured \emph{in vivo} \cite{ashikaga07:_transmural,ibrahim11:_tagging}
  or computed \emph{in silico} \cite{augustin16:_driver} --
  using constitutive material models which are derived from \emph{ex vivo} measurements \cite{guccione1995finite}
  and material parameters fitted to clinical data \cite{augustin16:_driver, xi2013estimation}.

  To evaluate the accuracy of Laplace analysis
  for estimating global wall stresses and mechanical power in the LV
  we employed four FE based EM LV models
  which have been previously fitted and validated with clinical data \cite{augustin16:_driver}
  under pre-treatment conditions.
  These models provide reliable strain data at a high spatio-temporal resolution
  from which wall stresses, biomechanical power and work in the LV can be determined
  at the best possible accuracy.
  Laplace analysis was applied to these \emph{in silico} models
  to estimate hoop stress and mechanical power over a cardiac cycle
  and compared to the global ground truth data based on FE analysis.

\section{Methods}
  \label{sec:methods}

\subsection{Patient Data}

Data from four AS patients with clinical indication for aortic valve treatment,
all preceding a cardiac magnetic resonance study, were used (Tab.~\ref{tab:patients}).
AS treatment indicators included valve area and/or systolic pressure drop across the valve.
The study was approved by the institutional Research Ethics Committee following the ethical
guidelines of the 1975 Declaration of Helsinki.
Written informed consent was obtained from the participants' guardians.

\begin{table}[ht]
	\setlength{\tabcolsep}{1.8mm}
	\centering
	\resizebox{\linewidth}{!}{
	\begin{tabular}{lccccccccccccccc}
		\toprule
		& Sex & Age & EDV & ESV & SV & EF & HR & $p_{\rm{dia}}$ & $p_{\rm{sys}}$ & MAP & $h$ & $\Delta p$ & HT & MVR  \\
	    & & [years] & [ml] & [ml] & [ml] & [\%] & [min$^{-1}]$ & [mmHg] & [mmHg] & [mmHg] & [mm] & [mmHg] & \\
		\midrule
		A & F & $63$ & $112.0$ &  $46.0$ & $66.00$ & $58.93$ & $53$ & $74$ & $126$ &  $91.33$ & $12.0/12.5$ & $95$ & No  & No   \\
		B & M & $73$ & $121.0$ &  $54.7$ & $66.32$ & $54.81$ & $81$ & $75$ & $134$ &  $94.67$ & $11.2/13.8$ & $62$ & No  & Mild \\
		C & M & $54$ & $118.2$ &  $42.2$ & $76.14$ & $64.42$ & $75$ & $71$ & $141$ &  $94.33$ & $16.0/18.2$ & $79$ & Yes & Mild \\
		D & M & $85$ & $172.0$ & $103.0$ & $69.00$ & $40.12$ & $68$ & $79$ & $144$ & $100.67$ & $14.0/15.2$ & $59$ & Yes & No   \\
		\bottomrule
        \end{tabular}}
	\caption{Pre-treatment AS patient characteristics from MRI and non-invasive cuff pressure recordings
		including end-diastolic volume (EDV), end-systolic volume (ESV),
		stroke volume (SV), ejection fraction (EF), heart rate (HR),
		diastolic and systolic cuff pressures ($p_{\rm{dia}}$ and $p_{\rm{sys}}$),
		mean arterial pressure (MAP), wall thickness at the LV equator measured in septum / lateral free wall ($h$),
		pressure drop across aortic valve ($\Delta p$), presence of hypertension (HT) and mitral valve regurgitation (MVR).}
	\label{tab:patients}
\end{table}

  \subsection{Biomechanical FE model}
    \label{sec:biomechfemodel}

    The ventricular myocardium was modeled as a non-linear, hyperelastic, nearly
    incompressible and anisotropic material with a layered organization of
    myocytes and fibres that is characterized by a right-handed orthonormal set of basis vectors
    \cite{guccione1995finite,holzapfel2009constitutive}.
    These basis vectors consist of the fiber axis $\vec{f}_0(\vx)$,
    which coincides with the prevailing orientation of the myocytes at location $\vx$,
    the sheet axis $\vec{s}_0(\vx)$, and the sheet-normal axis $\vec{n}_0(\vx)$.
    The mechanical deformation of the tissue is described by Cauchy's
    equation of motion under stationary equilibrium assumptions leading
    to a quasi-static boundary value problem: for a given pressure
    $p(t)$, find the unknown displacement $\vu$ such that
    \begin{alignat}{3}
      \label{equ:bvp}
      - \nabla \cdot \sigmatot(\vu,t) & = 0 &\quad& \text{ in } \Omega \\
      \sigmatot(\vu,t) \cdot \vn & = - p(t) \, \vn &\quad& \text{ on } \Gamma_N \notag \\
      \sigmatot(\vu,t) \cdot \vn & = 0 &\quad& \text{ on } \Gamma_H \notag \\
      \vu & = 0 &\quad& \text{ on } \Gamma_D \notag
    \end{alignat}
    holds for $t \in [0, T]$.
    By $\Omega$ we denote the deformed geometry and by $\Gamma = \partial \Omega$ we
    define its boundary with $\Gamma = \overline{\Gamma_D} \cup \overline{\Gamma_H} \cup \overline{\Gamma_N}$
    and $\abs{\Gamma_D} > 0$. The normal outward vector of $\Gamma$ is denoted
    by $\vn$. The total Cauchy stress tensor $\sigmatot$ refers to the sum of a passive
    stress tensor $\sigmapas$ and an active stress tensor $\sigmaact$. That is,
    $\sigmatot = \sigmapas + \sigmaact$ with
    \begin{align}
      \sigmapas & = J^{-1} \tensor{F}
                          \left( 2 \, \frac{\partial \Psi(\tensor{C})}{\partial \tensor{C}} \right)
                          \tensor{F}^{\top}, \label{equ:stress_pas} \\
      \sigmaact & = J^{-1} \vec{F}
                          \left( S_a (\vec{f}_0 \cdot \tensor{C} \vec{f}_0 )^{-1} \vec{f}_0 \otimes \mathbf{f}_0 \right)
                          \tensor{F}^{\top}, \label{equ:stress_act}
    \end{align}
    where $\tensor{F}$ is the deformation gradient,
    $\Psi$ is the strain energy function,
    $\vec{f}_0$ is fiber orientation in the reference configuration,
    $J=\operatorname{det} \tensor{F}$ is the Jacobian,
    $\tensor{C}=\tensor{F}^\top\tensor{F}$ is the right Cauchy--Green strain tensor
    and $S_a$ is the scalar active contractile stress generated by the myocytes
    acting along $\vec{f}_0$.

    The passive behavior of myocardial tissue was modeled
    using two material models,
    either the transversely-isotropic Guccione et al. model \cite{guccione1995finite},
    \begin{equation}\label{eq:guccioneStrainEnergy}
      \Psi_{\mathrm{Gu}}(\tensor{C}) = \frac{\kappa}{2} \left( \log\,J \right)^2 + \frac{a}{2}\left[\exp(\mathcal{Q})-1\right],
    \end{equation}
    where
    \begin{equation}\label{eq:guccioneQ}
      \mathcal{Q} =
        b_{\mathrm{f}} (\vec{f}_0\cdot\isotensor{E}\vec{f}_0)^2 +
        b_{\mathrm{t}} \left[(\vec{s}_0\cdot\isotensor{E}\vec{s}_0)^2+
                              (\vec{n}_0\cdot\isotensor{E}\vec{n}_0)^2+
                              2(\vec{s}_0\cdot\isotensor{E}\vec{n}_0)^2\right]+
        2b_{\mathrm{fs}} \left[(\vec{f}_0\cdot\isotensor{E}\vec{s}_0)^2+
                                 (\vec{f}_0\cdot\isotensor{E}\vec{n}_0)^2\right]
    \end{equation}
    and $\isotensor{E} = \frac{1}{2}(J^{-\frac{2}{3}}\tensor{C}-\tensor{I})$
    is the modified isochoric Green--Lagrange strain tensor,
    or the isotropic Demiray model \cite{Demiray1972}
    \begin{equation}\label{eq:demirayStrainEnergy}
      \Psi_{\mathrm{Dem}}(\tensor{C}) = \frac{\kappa}{2} \left( \log\,J \right)^2 + \frac{a}{2\,b}
                          \left\{ \exp \left[ b\, \big( \tr(\isotensor{C}) - 3 \big)\right] -1 \right\},
    \end{equation}
     with $\isotensor{C} = J^{-\frac{2}{3}}\tensor{C}$ the modified isochoric right Cauchy--Green tensor.
    In both models, Eqs.~\eqref{eq:guccioneStrainEnergy} and \eqref{eq:demirayStrainEnergy}, the bulk modulus
    $\kappa$, which serves as a penalty parameter to enforce near incompressibility, was chosen as $\kappa = 650$~kPa.

   A simplified phenomenological  contractile model \cite{niederer2011length} was used
   to represent active stress generation:
    \begin{equation} \label{eq:activeStressGeneration}
        S_{\rm a} = S_{\rm peak} \,
        \phi(\lambda) \,
        \operatorname{tanh}^2 \left( \frac{t_\mathrm{s}}{\tau_{\mathrm{c}}} \right) \,
        \operatorname{tanh}^2 \left( \frac{t_{\mathrm{dur}} - t_{\mathrm{s}}}{\tau_{\mathrm{r}}} \right)
        \qquad \text{for } 0< t_s < t_{\mathrm{dur}},
    \end{equation}
    where
    $S_{\mathrm{peak}}$ is the peak isometric tension,
    $\phi (\lambda)$ is a non-linear function dependent on fiber stretch
    $\lambda=\left|\tensor{F}\vec{f}_0\right|$ describing the length-dependence of active stress generation,
    $t_{\mathrm{s}}$ is the onset of contraction,
    $\tau_{\mathrm{c}}$ is the upstroke time constant,
    $t_{\mathrm{dur}}$ is the active stress transient duration and
    $\tau_{\mathrm{r}}$ is the downstroke time constant.
    This simplified model allows efficient fitting to patient data as the parameters for peak stress, $S_{\rm peak}$,
    and time constant of contraction, $\tau_{\mathrm{c}}$, and twitch duration, $t_{\mathrm{dur}}$,
    are related to the two clinical key parameters
    of interest, peak pressure and maximum rate of pressure increase, in an intuitive manner.

    Solving these equations under given mechanical boundary conditions
    using the FE method at a sufficiently high spatio-temporal discretization
    provides  an accurate description of tissue kinematics. 
    Computed displacement $\vu$ serve as input then in a post-processing procedure
    to evaluate wall stresses $\sigmatot(\vx,t)$ and to compute
    internal power expended by the LV (see Sec.~\ref{sec:intpowerandwork}).

    A Newton scheme was applied in each time step to linearize the nonlinear
    boundary value problem \eqref{equ:bvp} yielding a non-symmetric FE system.
    The linear FE system was solved by a parallel GMRES
    algorithm with an algebraic multigrid preconditioner. For the Newton scheme
    a relative tolerance of $1.0e^{-5}$ and an absolute tolerance of $1.0e^{-8}$
    was used as stopping criterion.
  \subsectionend{Biomechanical finite element model}


  \subsection{Verification of finite element model}
    \label{sec:verificationoffemodel}
\begin{figure}[!ht]
      \centering
      \subfloat[\;Geometrical setup.]{ \includegraphics[scale=0.7]{images2d_2} \label{fig:sph:geom} }
      \hspace{1.5cm}
      \subfloat[\;Spherical coordinate system.]{ \includegraphics[scale=0.7]{images3d_2} \label{fig:sph:csys} }
      \hspace{1.5cm}
      \subfloat[\;Mechanical boundary conditions.]{ \includegraphics[scale=0.7]{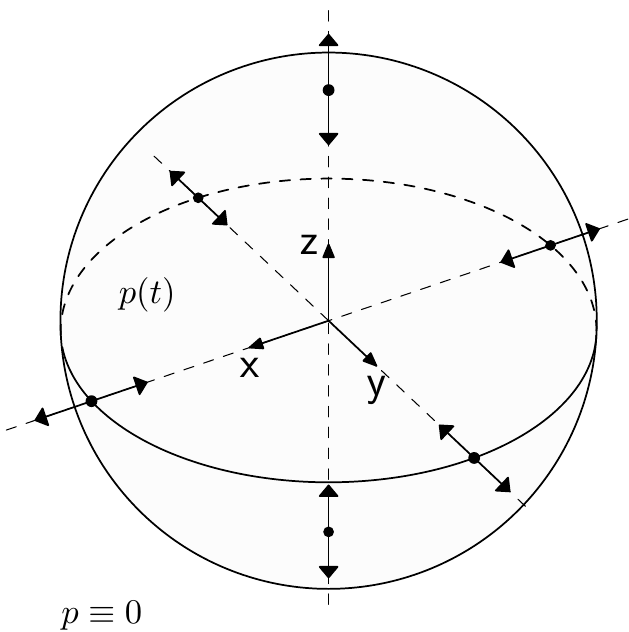} \label{fig:sph:bcs} }
      \caption{Geometric setup, spherical coordinate system and displacement boundary conditions.}
      \label{fig:sph}
    \end{figure}

    To verify the FE-based calculation of stress-derived metrics, a geometrically
    simple and well-studied benchmark problem was chosen
    for which circumferential hoop stresses can be found from Laplace's law
    under the following assumptions:
    \begin{itemize}
      \item[\textbf{(A1)}] The wall material is isotropic.
      \item[\textbf{(A2)}] The shape is a symmetric spherical shell with inner radius, $r$, and outer radius, $R$.
      \item[\textbf{(A3)}] The thickness of the wall, $h = R-r$, is sufficiently small, that is,
      the wall thickness to radius of curvature ratio is small, $h/r \ll 1$.
    \end{itemize}
    Since all these assumptions are violated in the LV
    which is orthotropic \textbf{(A1)}, non-spherical in shape \textbf{(A2)}, and thick-walled
    \textbf{(A3)} with $h/r \approx 1$, differences between Laplace analysis and FE computation are to be expected.
    Three configurations were considered, an ideal thin-walled spherical shell, $\sphh$,
    which complies with all assumptions \textbf{(A1)}--\textbf{(A3)}
    and thus can serve as a reference for FE validation,
    and two thicker-walled spheres, $\sphH$ and $\sphHH$,
    where assumption \textbf{(A3)} is increasingly violated.
    Geometries and mechanical boundary conditions are illustrated in Fig.~\subref*{fig:sph:geom}--\subref*{fig:sph:bcs}.

    The inner radius was chosen as $r=15.0$~mm in all models
    with $h$ varying from $h=0.5$~mm to $h=15.0$~mm in $\sphh$ and $\sphHH$, respectively.
    The choices for $\sphHH$
    are representative of the $h/r$ ratios found in the LVs of patients in this study (Tab.~\ref{tab:exp:setup:thin}).
    In line with assumption~\textbf{(A1)}, the nonlinear isotropic material law
    stated in Eq.~\eqref{eq:demirayStrainEnergy}
    was employed with $a = 10$~kPa and $b = 8$.
    Passive inflation experiments were performed by solving \eqref{equ:bvp} with $\sigmaact \equiv 0$
    and applying a pressure $p$ in the range from $0$ to $4$~kPa to the endocardial surface, $\Gamma_{\mathrm{endo}}$,
    which covers the range of pressures observed \emph{in vivo} during diastole.
    Pressure at the epicardial surface, $\Gamma_{\mathrm{epi}}$, was assumed to be zero.
    To render the solution of this pure Neumann problem unique, displacement boundary conditions were enforced
    at the intersections of the Cartesian axes with the epicardial surface
    by restricting displacements to the respective intersecting axes (see Fig~\subref*{fig:sph:bcs}).
    Unstructured tetrahedral FE meshes were generated for the $\sphh$, $\sphH$  and $\sphHH$ geometries
    where the mean spatial resolution, $\bar{\dx}$, was increased
    until solutions were deemed converged.
%
%

%
    \begin{center}
      \begin{table}[h!]
        \centering
	\resizebox{\linewidth}{!}{
        \begin{tabular*}{500pt}{@{\extracolsep\fill}l|cccc@{\extracolsep\fill}}
          \toprule
          & $r$ [mm] &  $R$ [mm] & $h$ [mm] & $h/r$ \\
          \midrule
          $\sphh / \sphH / \sphHH$  & $15/15/15$ & $15.5/17.5/30.0$ & $0.5/2.5/15.0$ & $0.033/0.166/1.0$ \\
          \midrule
          $\lva$/ $\lvb$/ $\lvc$/ $\lvd$ & $16.9/19.5/17.1/22.1$ & $30.6/33.6/36.9/38.5$ & $13.7/14.1/19.8/16.4$ & $0.82/0.72/1.16/0.74$ \\
          \bottomrule
      \end{tabular*}}
        \caption{Geometric parameters inner radius, $r$, outer radius, $R$, wall thickness, $h$ and wall thickness to radius of curvature ratio, $h/r$
         of spherical shell models $\sphh$, $\sphH$ and $\sphHH$ and of image-based anatomical LV models in the stress-free reference configuration. }
        \label{tab:exp:setup:thin}
      \end{table}
    \end{center}

\vspace{-0.8cm}


\subsection{LV model}

\subsubsection{Anatomical Modeling}
\begin{figure}[!ht]
	\centering
	\includegraphics[width=0.9\textwidth]{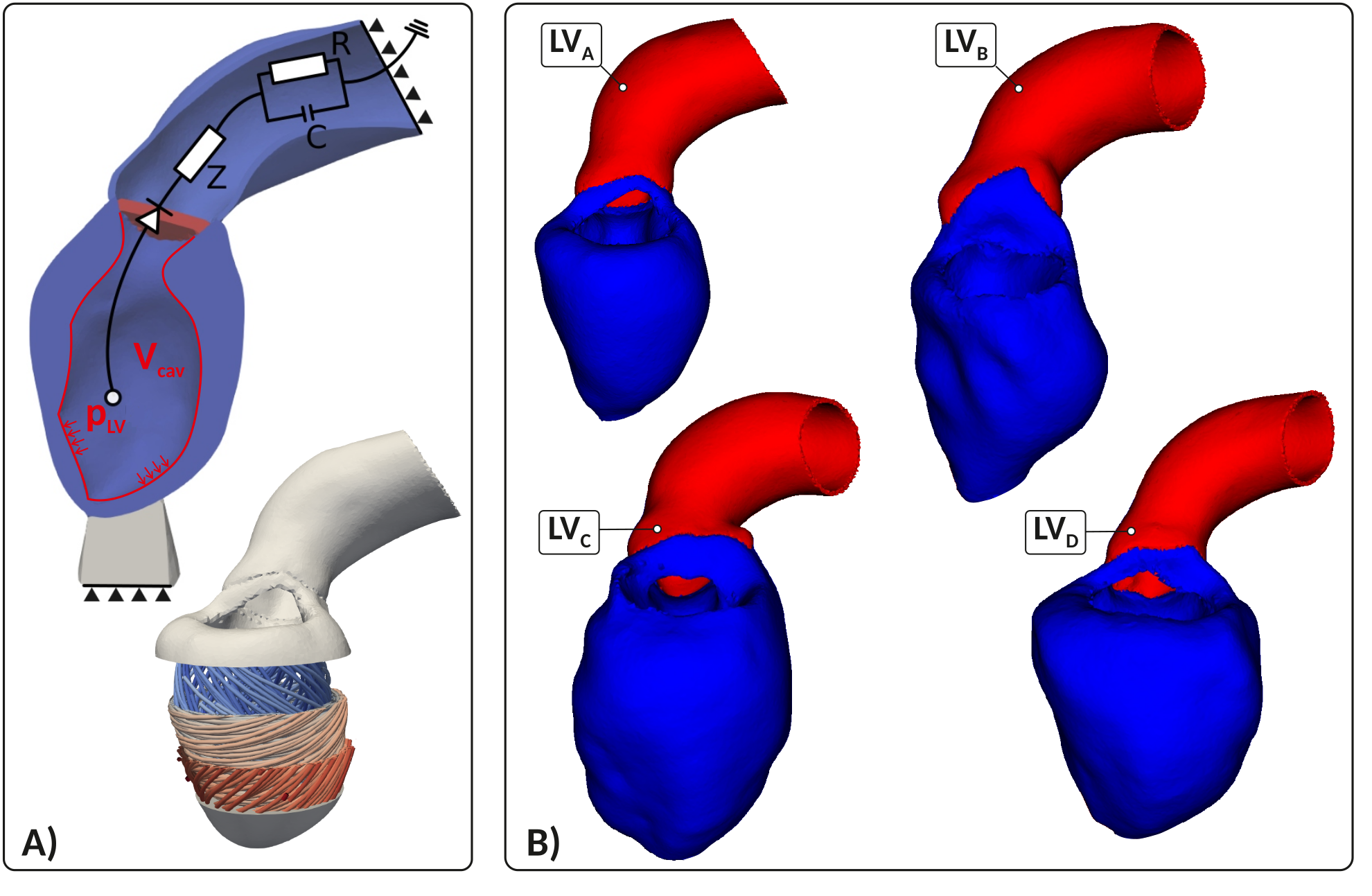}
	\caption{Image-based patient-specific LV anatomy models.
		(A) Shown are the FE model setup with Dirichlet (solid triangles)
		and Neumann boundary conditions controlled by a three-element Windkessel model of afterload,
		and fiber architecture (bottom panel).
		(B) Patient-specific anatomical models $\lva$, $\lvb$, $\lvc$ and $\lvd$ of LV and aorta
		constructed from a 3DWH MRI scan in end-diastolic configuration.}
	\label{fig:lv}
\end{figure}

FE meshes of the LV anatomy and aortic root were generated from 3D whole heart MRI
acquired at end diastole (ED) with $1.458 \times 1.548 \times 2$~mm resolution
at the German Heart Center Berlin.
Multi-label segmentation of the LV myocardium, LV cavity and aortic lumen was done
using the ZIB Amira software (\href{https://amira.zib.de/}{https://amira.zib.de/}).
Segmentations were smoothed and upsampled to a $0.1$~mm isotropic resolution \cite{crozier2015image}.
The wall of the aorta was automatically generated by dilation of the aortic lumen
with a thickness of $1.2$~mm,
and the aorta was clipped before the branch of the brachiocephalic artery.
Due to limited resolution valves were not segmented,
but were included in the FE model as a thin layer of tissue
for applying pressure boundary conditions and computation of cavity volume.
The multi-label segmentations were meshed using
CGAL (\href{http://www.cgal.org/}{http://www.cgal.org/})
with a target discretization of $1.25$~mm in the LV myocardium and $1$~mm in the aortic wall.
For the transversely-isotropic Guccione material model, see Eq.~(\ref{eq:guccioneStrainEnergy}), we equipped all
models with a rule-based fiber architecture \cite{bayer12:_ldrb},
where fibers rotated linearly from $-75^\circ$ at the epicardium to $+75^\circ$ at the endocardium (Fig.~\ref{fig:lv}.A).
All anatomical models built are shown in Fig.~\ref{fig:lv}.

\vspace{-0.8cm}

\subsubsection{Model fitting}
\label{sec:fittingoffelvmodel}
To remove rigid body motion and provide physiological boundary conditions
that allow a vertical movement of the LV base, as observed \emph{in-vivo},
mechanical boundary conditions were applied
by fixing the terminal rim of the clipped aorta (Fig.~\ref{fig:lv}.A)
and resting the apex of the LV on an elastic cushion
which was rigidly anchored at its base.
Constitutive relations were represented by Eq.~\eqref{eq:guccioneStrainEnergy}.
Using the ED geometry, default material parameters and an estimated ED pressure (EDP),
an initial guess of the stress-free reference configuration was computed
by unloading the model using a backward displacement method \cite{sellier11:_iterative,bols13:_unloading}.
Since clinically recorded data of the ED PV-relation (EDPVR) are often limited,
the Klotz relation  \cite{klotz2007computational} providing an empiric description of EDPVR,  $p(\volcav)$,
was used as target to steer the fitting of constitutive parameters.
In absence of accurate measurements of EDP
we refrained from fitting all material parameters to $p(\volcav)$.
Rather, default values for the parameters $b_f=18.48$, $b_t=3.58$ and $b_{fs}=1.627$ were used
as reported in the literature \cite{guccione1995finite},
and only the scaling parameter $a$ was adjusted individually for each patient.
With a given data point (EDV,EDP) $a$ was fitted to minimize the difference
in stress-free residual volume, $V_{\rm{0,dia}}$, between model and Klotz curve.
This yielded values for $a$ of $0.5$, $0.65$, $0.5$ and $0.5$ for the cases $\lva$, $\lvb$, $\lvc$ and $\lvd$, respectively.

\begin{figure}[!ht]
	\centering
	\includegraphics[width=0.4\textwidth]{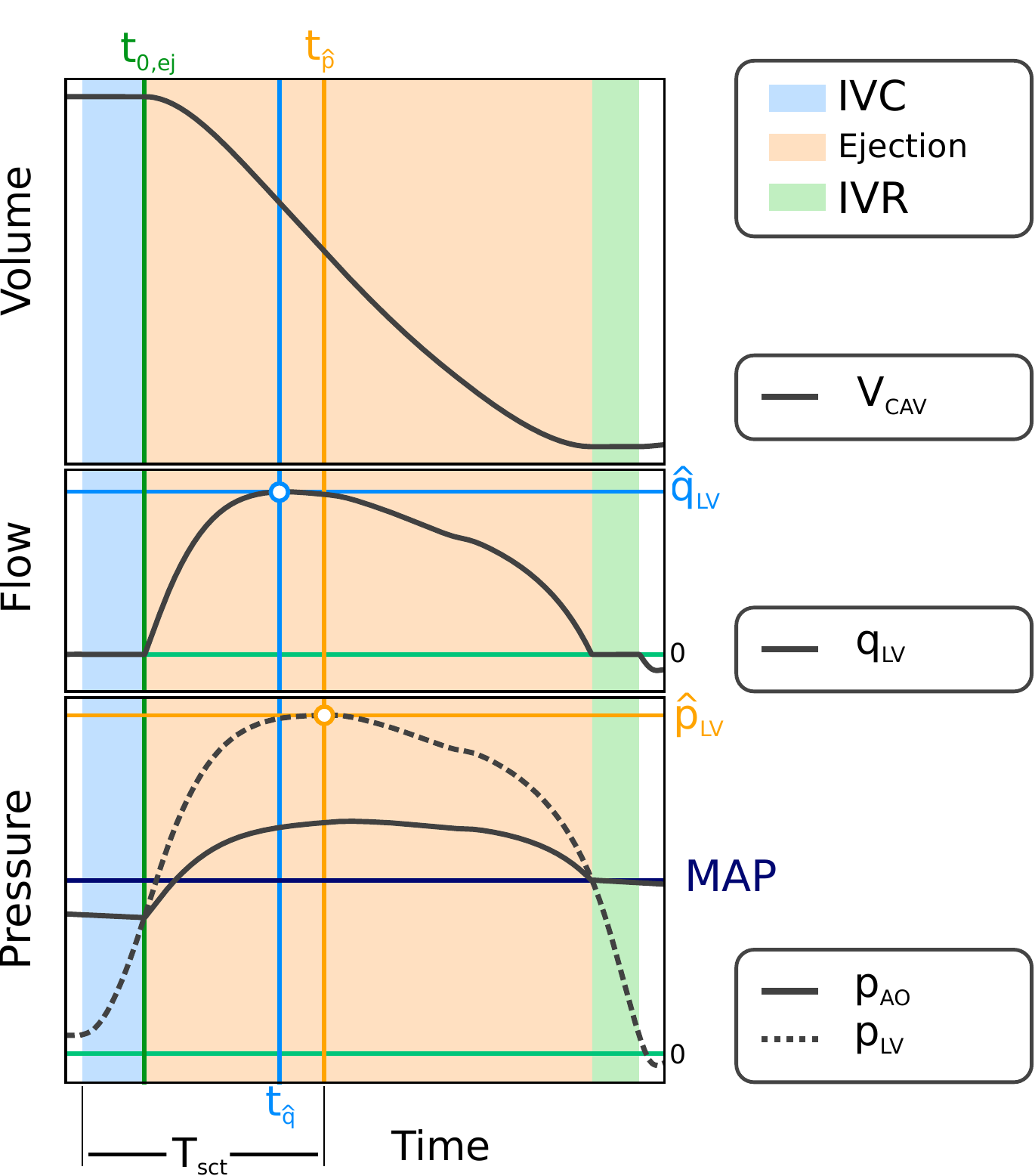}
	\caption{Fitting of the afterload model. Measured input data comprise $\volcav(t)$ (top panel),
	derived flow $q_{\mathrm{LV}} = d \volcav/dt$ (mid panel),
	$\plvpeak$ and pressure drop $\Delta p = \plvpeak - \paopeak$
	along with simulated pressure traces $\plv$ and $\pao$ (bottom panel),
	with annotations of onset of ejection, $\tejs$, instant of peak pressure in the LV, $\tplvpeak$,
	cardiac contraction time, $\Tsct$, and the instant of peak flow, $\tqlvpeak$.}
	\label{fig:pv_defs}
\end{figure}

A three-element Windkessel model of LV afterload was used
to provide the pressure-flow relationship during ejection \cite{westerhof71:_wk3} (see Fig.~\ref{fig:pv_defs}).
LV models were parameterized to match clinically recorded PV-data
using LV cavity volume traces, $\volcav(t)$, determined from Cine-MRI
with a temporal resolution of $45.28$, $29.63$, $32.00$ and $35.29$ ms for
$\lva$, $\lvb$, $\lvc$ and $\lvd$, respectively.
Continuously monitored invasive pressure recordings were not available
as catheterization was not indicated.
Peak pressure in the LV was determined
by estimating peak pressure in the aortic root from cuff pressure measurements
and by determining the pressure drop at peak flow across the aortic valve
from ultrasound flow measurements using Bernoulli's law \cite{donati17:_bernoulli}.
Windkessel parameters representing the aortic input impedance, $Z$,
comprising the flow resistance of aortic valve, $Z_{v}$,
and the characteristic input impedance of the aorta, $Z_{\rm c}$,
as well as resistance $R$ and compliance $C$ of the arterial system
were fit to reproduce estimated LV peak pressure
using measured volume traces $\volcav(t)$ as input.

In a final step, active mechanical properties were fit
to the same hemodynamic data used for fitting the afterload model.
A reaction-eikonal model was used to generate activation sequences
and simulate action potential propagation in the LV \cite{neic17:_efficient}.
Active stress generation was triggered with a prescribed electromechanical delay
when the upstroke of the action potential crossed the $-40$~mV threshold.
Parameters peak stress, $S_{\rm peak}$, time constant of contraction, $\tau_{\mathrm{c}}$,
and twitch duration, $t_{\mathrm{dur}}$, were adjusted manually to fit
peak pressure, $\plvpeak$, duration of pressure pulse and flow.
Due to the intuitive link of the active stress model given in Eq.~\eqref{eq:activeStressGeneration}
with the fitting targets, a satisfactory fit was achieved within $\le$ 5 simulation runs.
The goodness of fit was deemed sufficiently accurate when the clinically measured metrics
EF, SV, MAP and peak LV pressure, $\plvpeak$,
were matched within a margin of error of $\pm 5\%$.
Clinical input data and fitted model parameters are summarized in Tab.~\ref{tab:model_fitted_params}.
    \begin{center}
    	\begin{table}[h!]
    		\centering
	\resizebox{\linewidth}{!}{
    		\begin{tabular*}{500pt}{@{\extracolsep\fill}l|cccc|cc|c@{\extracolsep\fill}}
    			\toprule
    			& \multicolumn{4}{c}{afterload} & \multicolumn{2}{c}{active stress} & passive mechanical model \\
    			& $Z_{v} \left[\txtfrac{kPa.ml}{ms}\right]$ & $Z_{a} \left[\txtfrac{kPa.ml}{ms}\right]$ & $R \left[\txtfrac{kPa.ml}{ms}\right]$ & $C \left[\txtfrac{ml}{kPa}\right]$
    			& $S_{\rm peak}$ [kPa] & $\tau_{c}$ [ms] & $a$ [kPa] \\
    			\midrule
          $\lva$ & $35.82$ & $26.00$ & $187.74$ & $15.23$ & $69$ & $80$ &  $0.5$ \\
          $\lvb$ & $16.08$ & $11.03$ &  $72.65$ & $26.62$ & $85$ & $35$ & $0.65$ \\
          $\lvc$ & $15.93$ & $12.78$ &  $77.34$ & $25.50$ & $63$ & $40$ &  $0.5$ \\
          $\lvd$ & $22.34$ & $11.09$ &  $62.73$ & $30.93$ & $98$ & $58$ &  $0.5$ \\
    			\bottomrule
                \end{tabular*}}
    		\caption{Fitted parameters of circulatory, active stress and passive mechanical
    			model components.}
    		\label{tab:model_fitted_params}
    	\end{table}
    \end{center}


\vspace{-1cm}
  \subsection{Myocardial wall stresses}
    \label{sec:myowallstresses}

    Stresses can be computed from deformations $\vu$ using constitutive
    material models based on \emph{ex vivo} experimental data which link stresses to strains.
    In the FE model stress tensors $\sigmatot(\mathbf{x},t)$ are computed
    by evaluating Eqs.~\eqref{equ:stress_pas} and \eqref{equ:stress_act},
    which yields a $3 \times 3$ tensor where only 6 components are independent for symmetry reasons.
    For the models $\sphh$, $\sphH$ and $\sphHH$ the stress tensor simplifies.
    Due to the assumption of isotropy in \textbf{(A1)} and symmetry in \textbf{(A2)},
    any solution, if expressed in a spherical coordinate system, must also be symmetric.
    Quantities computed in the FE Cartesian coordinate system are recast in spherical coordinates
    as defined in Fig.~\subref*{fig:sph:csys} using a projection matrix, $\mathbf{P}$.
    For the total Cauchy stress tensor $\sigmatot$ we obtain
    \begin{equation}
      \label{equ:sig:sph}
      \sigmatot = \left(
      \begin{array}{ccc}
        \sigma_{xx} & \sigma_{xy} & \sigma_{xz} \\
        \sigma_{yx} & \sigma_{yy} & \sigma_{yz} \\
        \sigma_{zx} & \sigma_{zy} & \sigma_{zz}
      \end{array}
      \right) = \mathbf{P} \left(
      \begin{array}{ccc}
        \sigma_{r r} & \sigma_{r \varphi} & \sigma_{r \theta} \\
        \sigma_{\varphi r} & \sigma_{\varphi \varphi} & \sigma_{\varphi \theta} \\
        \sigma_{\theta r} & \sigma_{\theta \varphi} & \sigma_{\theta \theta} \\
      \end{array}
      \right) \mathbf{P}^{\top} = \mathbf{P} \, \sigmatotsph \, \mathbf{P}^{\top}
    \end{equation}
    with the projection matrix $\mathbf{P} = ( \ve_{r}, \ve_{\varphi},  \ve_{\theta} )$
    and $r \in \R^{+}, \theta \in [0, \pi], \varphi \in [0, 2 \, \pi)$.

    While all quantities in the spherical models must be perfectly symmetric this is not
    necessarily the case in the FE solutions.
    Depending on spatial resolution and boundary conditions, a minor numerical jitter around mean values will inevitably occur.
    For comparing FE with Laplace analysis averaged mean quantities were therefore computed over the entire domain by
    \begin{equation}
      \label{equ:avg:stress}
      \subscript{\overline{\sigma}}{\star\star}(\vu, t) = \frac{1}{\abs{\Omega}} \, \int \limits_{\Omega} \sigma_{\star\star}(\vu, t) \dvx
    \end{equation}
    with $\vu$ being the FE solution at time $t$ and $\star \in \{ r, \theta, \varphi \}$.

    In a thin-walled spherical shell stresses in azimuthal and meridional direction
    must be equal due to symmetry, that is,
    $\sigmacirc = \sigma_{\varphi \varphi} = \sigma_{\theta \theta}$,
    and with $h \ll r$ radial stresses can be assumed to be negligible
    relative to circumferential hoop stresses, that is, $0 \approx \sigma_{rr} \ll \sigmacirc$.
    The stress tensor in spherical coordinates simplifies therefore to
    \begin{equation}
      \label{equ:sig:sph:smp}
      \sigmatotsph \approx \left(
      \begin{array}{ccc}
        0 & 0 & 0 \\
        0 & \sigmacirc & 0 \\
        0 & 0 & \sigmacirc \\
      \end{array}
      \right). 
    \end{equation}

    As a reference for verifying FE-based stresses different variants of Laplace's law were used.
    In particular, we use an extension of Laplace's law
    which takes into account the finite thickness of the wall
    \begin{equation}
      \label{equ:lol:ext}
      \sigmacirc = \frac{p \, r}{2 \, h \left( 1 + \frac{h}{2 \, r} \right) } = \sigmaLext
    \end{equation}
    and refer to stress estimates based on this formula as Laplace stress in thick-walled spherical shells, $\sigmaLext$.
    Exploiting assumption~\textbf{(A3)}, that is, $\frac{h}{r} \ll 1$ and thus
    $\left(1+\frac{h}{2 \, r}\right) \approx 1$, allows a further simplification yielding
    \begin{equation}
      \label{equ:lol:smp}
      \sigmacirc =  \frac{p \, r}{2 \, h} = \sigmaLsmp
    \end{equation}
    which we refer to as Laplace stress in thin-walled spherical shells, $\sigmaLsmp$.
    Finally, we consider a volume-based estimation of $\sigmacirc$ \cite{mirsky73:_assessment}
    defined as
    \begin{equation}
      \label{equ:lol:vol}
      \sigmacirc = \frac{p}{\left( \frac{\volcav + \volmyo}{\volcav} \right)^{2/3}-1} = \sigmaLvol,
    \end{equation}
    which has been used previously in clinical studies \cite{alter16:_wallstress},
    We note that Eq.~\eqref{equ:lol:vol} is equivalent to Eq.~\eqref{equ:lol:ext}
    for a spherical shell.
    However, when applied to a non-spherical structure such as the LV this is not the case.
    Eq.~\eqref{equ:lol:vol} may offer advantages
    as the determination of $\volcav$ and $\volmyo$
    may be less ambiguous
    than the determination of a representative inner radius $r$ and wall thickness $h$
    (see Sec.~\ref{sec:determine_clinical_input}).



  \subsection{Myocardial power and work}
    \label{sec:intpowerandwork}

    For a given displacement $\vu$ at time $t \in [0,T]$,
    where $T$ refers to the duration of a cardiac cycle
    and $t_0 = \tED = 0$ marks the end of diastole,
    the biomechanical power density, $p_{\mathrm{int}}$,
    generated or consumed at location $\mathbf{x}$ within the LV wall
    can be computed by evaluating
    \begin{equation}
      \label{equ:ihp_density}
      p_{\mathrm{int}}(\vx,t) = \sigmatot(\vu, t) : \straindot(\vu, t),
    \end{equation}
    where $\straindot$ is the strain rate tensor
    and $\tensor{A}:\tensor{B} = \tr(\tensor{A}^{\top}\,\tensor{B})$ denotes the double contraction of two tensors, see e.g. \cite{Holzapfel2000}
    for further details. Integration of Eq.~\eqref{equ:ihp_density} over the entire
    myocardial wall yields the global biomechanical power, $\powerint$,
    \begin{equation}
      \label{equ:ipower}
      \powerint(t) = \int \limits_{\Omega} \sigmatot(\vu, t) : \straindot (\vu, t) \dvx
    \end{equation}
    and integration of Eq.~\eqref{equ:ipower} over time
    yields an expression of biomechanical work, $\workint$, performed
    \begin{equation}
      \label{equ:iwork}
      \workint = \int \limits_{t_0}^{t} \powerint(\tau) d\tau.
    \end{equation}

    Based on Laplace's law biomechanical power can be estimated using
    \begin{equation}
      \label{equ:lol:ipow}
      \powerintstar(t) = \volmyo(t) \, \sigmaLstar(t) \, \underbrace{\left( \frac{\dot{r}(t)}{r_0} + \frac{\dot{R}(t)}{R_0} \right)}_{\approx \straindot_{\mathrm{circ}} }
    \end{equation}
    where $\star$ denotes which formula was used for estimating the circumferential wall stress
    that is, $\star \in \left\{ \text{h}, \text{H}, \text{V} \right\}$
    and $\straindot_{\mathrm{circ}}$ approximates circumferential strains.
    For a derivation of Eq.~\eqref{equ:lol:ipow} see supplementary material. 
    Laplace-based mechanical work is estimated analogously to Eq.~\eqref{equ:iwork}
    yielding
    \begin{equation}
    \label{equ:lol:iwork}
    \workintstar(t) = \int \limits_{t_0}^{t} \powerintstar(\tau) \intd{\tau}.
    \end{equation}

     In addition, a recently introduced also Laplace-based relative power indicator, $\ihp$,
     was evaluated
     which attempts to estimate the power generated by the LV around the instant of peak pressure, \tplvpeak.
     Based on \cite{lippincott12:_illustrated},
     the mechanical work expended, $\ihw$, during contraction time, $\Tsct$,
     defined as the time elapsed between the onset of isovolumetric contraction at $\tED$
     and the instant of peak stress in the LV at $\tplvpeak$,
     is approximated by
     \begin{equation}
       \ihw = \volmyo \, \sigmaLstar.
       \label{equ:ihw:clin}
     \end{equation}
    $\ihw$ is interpreted as a measure of the mechanical potential energy stored in the LV
    from which a measure of the peak biomechanical power
    generated by the LV between $\tED$ and $\tplvpeak$ is derived then by
    \begin{equation}
      \label{equ:ihp:clin}
      \ihp = \frac{\ihw}{\Tsct}. 
    \end{equation}
    Note that Eq.~\eqref{equ:ihp:clin}, in contrast to Eq.~\eqref{equ:lol:ipow},
    does not include any measure of \straindot.
    Thus, while consistent in terms of physical units, $\ihp$ must be considered
    a relative indicator and not a physical measure of power.


  \subsection{Hydrodynamic power and work}
    \label{sec:extpvwork}

    Hydrodynamic power, $\powerext$, is given by
    \begin{equation}
      \label{equ:epower}
      \powerext =  p \, q = p \, \diff{\volcav}{t}
    \end{equation}
    where $p$ is the hydrostatic pressure acting at endocardial surface,
    $\Gamma_{\mathrm{endo}}$
    and $q$ represents blood flow out of the LV cavity during ejection.
    Hydrodynamic work, $\workext$,
    is then the work expended by the LV myocardium when changing the volume of its cavity,
    $\volcav$,
    given by
     \begin{equation}
      \label{equ:ework_pq}
      \workext = \int \limits_{t_0}^{t} p(\tau) \, q(\tau) \intd{\tau},
    \end{equation}
    or, equivalently, expressed as PV work by
    \begin{equation}
      \label{equ:ework_pv}
      \workext = \int \limits_{\volcav_0}^{\volcav_1} p(\volcav) \intd{\volcav}.
    \end{equation}

    In absence of  active stresses, i.e. $\sigmaact = 0$,
    and isovolumetric contstraints imposed by valves upon $\volcav$,
    $\powerint \equiv \powerext$ must hold.
    Under such conditions, external work can therefore serve as a reference
    for validating the FE-based computation of internal power and work.
    This is not necessarily the case during the isovolumetric phases of a heartbeat
    where internal work may be expended which does not necessarily manifest as  external work.
    During these phases, changes in $\sigmatot$ occur which may entail shape changes of the LV myocardium
    and thus induce a non-zero strain rate tensor $\straindot$.
    However, due to the isovlumetric constraints imposed by the incompressibility of the blood pool
    and the closed state of all valves, no global change in cavity volume can occur, i.e.\ $\mathrm{d}\volcav=0$.

   Under healthy conditions, hydrodynamic power in the LV cavity equals the power delivered to the arterial system as transvalvular power losses are small.
   However, in AS cases where transvalvular pressure gradients, $\Delta p$, are significant,
   the effective hydrodynamic power externally delivered to the arterial system is reduced.
   Following \cite{fernandes17:_beyond}, we define external hydrodynamic heart power, $\ehp$, as
   \begin{equation}
      \label{equ:ehp:clin}
      \ehp = \powerextmeanao = \frac{1}{\Tsys} \int \limits_{\tED}^{\tES} \pao \, q \intd{t} \approx \MAP \cdot \CO,
   \end{equation}
   where $\MAP$ and $\CO$ are mean arterial pressure and cardiac output, respectively,
   and $\pao = \plv - \Delta p$ is the pressure in the aorta ascendens.
   Power efficiency, $\powereff$, has been defined previously in \cite{fernandes17:_beyond} as the ratio
   \begin{equation}
        \label{equ:powereff:clin}
        \powereffclin = \frac{\ehp}{\ihp}.
    \end{equation}
    Since $\powereff$ essentially relates the mean hydrodynamic power delivered to the arterial system
    to the peak biomechanical power generated by the LV myocardium during systole,
    $\powereff$ can be expressed as
      \begin{equation}
        \label{equ:powereff}
        \powereff = \frac{\powerextmeanao}{\powerintpeak} \approx  \frac{\powerextmeanao}{\powerextpeak}.
    \end{equation}
    where  $\powerintpeak$, $\powerextpeak$ and $\powerextmeanao$
    are determined based on Eqs.~\eqref{equ:ipower}, \eqref{equ:epower}
    and \eqref{equ:ehp:clin}, respectively.
    That is, $\powereff$ can be estimated from hemodynamic data using $\powerextpeak$
    or from LV deformation analysis using $\powerintpeak$.
    Unlike $\powereffclin$ in Eq.~\eqref{equ:powereff:clin},
    which compares an absolute measure of external hydrodynamic power
    to a relative indicator of internal biomechanical peak power,
    Eq.~\eqref{equ:powereff} provides a physically consistent comparison.

  \subsectionend{External pressure-volume work}


  \subsection{Evaluation of Laplace-based assessment of wall stress and power}
    \label{sec:evalofclinwallstress}


    Human EM LV models which were validated against clinical data
   (see Sec.~\ref{sec:fittingoffelvmodel}),
    provided accurate ground truth data on strains $\strain(\vu, t)$
    and stresses $\sigmatot(\vu, t)$ in the LV wall.
    Using these as reference, Laplace analysis was applied to the \emph{in silico} models
    to assess its accuracy and validity.

  \subsubsection{Determination of clinical input data for Laplace-based analysis}
    \label{sec:determine_clinical_input}

    Geometric input parameters $r$ and $h$
    required for Laplace analysis  must be determined from clinical imaging datasets.
     As LV shape deviates markedly from a spherical shell
     representative mean parameters of $r$ and $h$ must be found.
     Since there is no unique best solution to establish a geometric correspondence
     between LV shape and a spherical shell,
     various methods have been used in clinical applications.
     Typically, transverse slices from short-axis Cine-MRI scans were analyzed
     to measure, either manually or semi-automatically, $r$ and $h$,
     where $h$ is measured in the postero-lateral wall, the septum or an average is taken.
     The analysis is either carried out in one representative mid-cavity LV short axis slice,
     or a number of slices is selected
     to capture representative basal, mid-cavity and apical LV cross-sections.

\begin{figure}[!ht]
      \centering
      \subfloat[\;$z$-slice selection.]{ \includegraphics[scale=0.325]{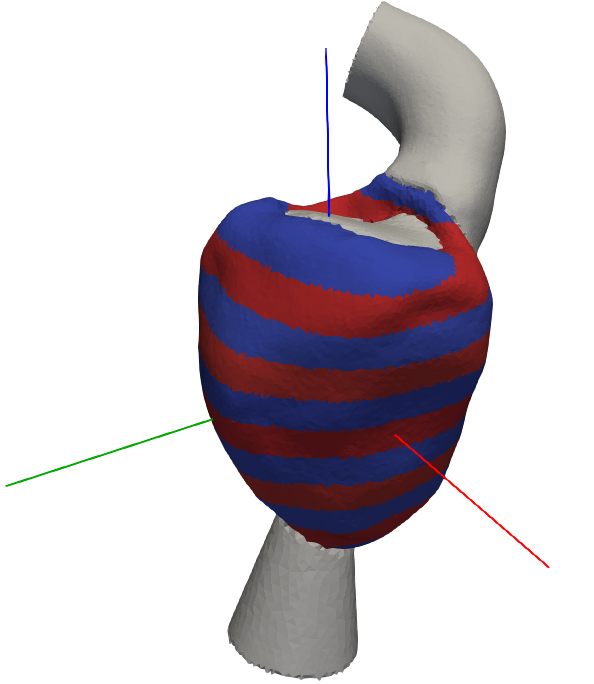} \label{fig:lv:slicing} } \hfill
      \subfloat[\;Determination of $r$ and $h$.]{ \includegraphics[scale=0.325]{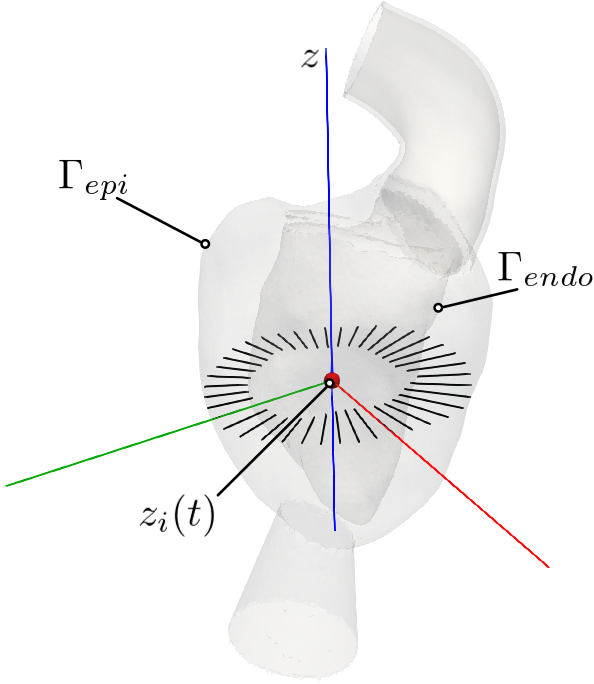} \label{fig:lv:slice_avg} } \hfill
      \subfloat[\;Time courses $\bar{r}(t)$ and $\bar{h}(t)$.]{ \includegraphics[scale=0.65]{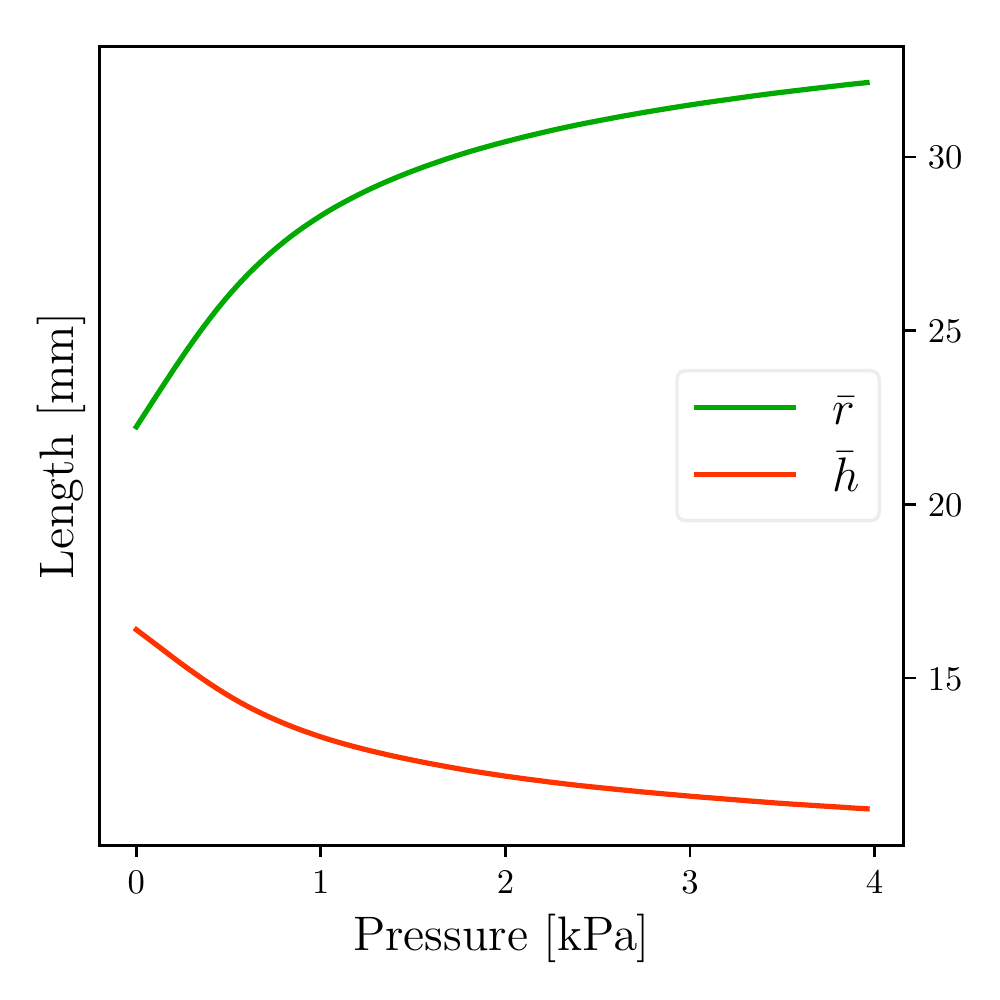} \label{fig:lv:mean_r_h_of_t} } \hfill
      \caption{Determination of input parameters radius $\bar{r}$ and wall width $\bar{h}$
      	for Laplace analysis.
      	(A) Automated LV slicing along long-axis $\mathbf{z}$.
      	(B) Sampling of variations in $r$, $R$ and $h$ within a slice, $z_i$.
      	(C) Averaged parameters $\bar{r}$ and $\bar{h}$ as a function of $p$. }
      \label{fig:lv:rad_determination}
    \end{figure}

     Similar issues arise when applying Laplace analysis to \emph{in silico} datasets.
     In order to extract  $r$ and $h$ as objectively as possible without operator bias,
     automated processing workflows were implemented (see Fig.~\ref{fig:lv:rad_determination}).
     Analogous to the $z$-slice selection in MRI protocols,
     the unstructured FE meshes of  the LV models were decomposed
     into slices of $\approx 8$~mm resolution,
     comparable to the MRI out-of-plane resolution.
     Decomposition was achieved by first determining the long axis, $\mathbf{z}$, of the LV
     using principal component analysis (see Fig.~\subref*{fig:lv:slicing}),
     which yielded, depending on the spatial extent of the LV long-axis of a given model,
     between $10$ and $14$ slices, each slice $i$ is centered around $z_i$.
     A mean $z$ coordinate of the LV in its current configuration, $\bar{z}_{c}(t)$, was computed
     to define the center slice plane using the long-axis unit vector, $\mathbf{e_{\mathrm{z}}}$,
     and the center $z_i(t)$ of individual slices was shifted,
     keeping slice width and distance to the LV center $\bar{z}_{c}(t)$ constant.
     Within a selected plane, radial vectors, $\mathbf{r}_{i,j}(t)$, were computed
     that emanated from $z_i$ and were oriented in polar angles $\varphi_j$
     ranging from $0$ to $360^\circ$ with an angular sampling of $\Delta \varphi=9^\circ$.
     For each vector $\mathbf{r}_{i,j}(t)$, the intersection with surfaces,
     $\Gamma_{\mathrm{endo}}(t)$ and $\Gamma_{\mathrm{epi}}(t)$, was determined,
     yielding $N=2\pi/\Delta \varphi$ inner radii, $r_{i,j}(t)$, outer radii, $R_{i,j}(t)$ and wall widths, $h_{i,j}(t)=R_{i,j}(t)-r_{i,j}(t)$
     (Fig.~\subref*{fig:lv:slice_avg}).
     Mean radius, $\bar{r}_i(t)$, and wall width, $\bar{h}_i(t)$, were determined
     as the arithmetic average
     \begin{equation}
        \bar{r}_i(t) = \frac{1}{N}\sum_{j=1}^{N}  r_{i,j}(t) \qquad \text{and} \qquad \bar{h}_i(t) = \frac{1}{N} \sum_{j=1}^{N}  h_{i,j}(t).
     \end{equation}
     Finally, multi-slice mean $\bar{r}$ and $\bar{h}$ were computed by averaging over $M$ slices
     \begin{equation}
        \bar{r}(t) = \frac{1}{M}\sum_{i=1}^{M} \bar{r}_{i}(t) \qquad \text{and} \qquad \bar{h}(t) = \frac{1}{M} \sum_{i=1}^{M} \bar{h}_{i}(t). \label{equ:_mean_r_h}
     \end{equation}
    The time course of the mean values $\bar{r}(t)$ and $\bar{h}(t)$ (see Fig.~\subref*{fig:lv:mean_r_h_of_t})
    was plugged then into the respective Laplace equations
    to compute stress in Eqs.~\eqref{equ:lol:ext}, \eqref{equ:lol:smp},  and \eqref{equ:lol:vol},
    power in Eq.~ \eqref{equ:lol:ipow} and work in Eq.~\eqref{equ:lol:iwork}.

  \subsubsection{Simulation protocols and data analysis}
    \label{sec:_analyze_passive_inflation}
    To evaluate the influence of violating the assumptions \textbf{(A1)} and  \textbf{(A2)},
    passive inflation experiments were performed with LV models and the anisotropic material given in Eq.~\eqref{eq:guccioneStrainEnergy}
    following the same protocol as applied before to the spherical shell models
    $\sphh$, $\sphH$ and $\sphHH$ in Sec.~\ref{sec:verificationoffemodel}.
    Laplace-based stress estimates $\sigmaLsmp$, $\sigmaLext$ and $\sigmaLvol$
    were compared to the mean stresses obtained from the FE solution.
    Stresses were evaluated with respect to an ellipsoidal coordinate system
    to facilitate a comparison with stresses in the spherical shell models
    (see Fig.~\subref*{fig:sph:csys}).
    An ellipsoidal coordinate system was constructed for the LV models
    by assigning fiber and sheet orientations using a rule-based method
    with a constant fiber angle of $0^\circ$ \cite{bayer12:_ldrb}.
    Stress components $\sigmarad(\vx)$, $\sigmaphi(\vx)$ and $\sigmatht(\vx)$ were averaged
    according to Eq.~\eqref{equ:avg:stress} yielding $\msigmarad$, $\msigmaphi$ and $\msigmatht$, respectively.
    Laplace-based estimation of power, $\powerintstar$,
    was compared to FE-based power, $\powerint$,
    and to external hydrodynamic power in the LV cavity, $\powerext$.


    Laplace analysis was applied to clinically fitted EM LV models $\lva$--$\lvd$
    to compare LV stress $\sigmaLstar$, power $\powerintstar$ and $\ihp$
    over an entire systolic cycle
    to the FE-based stresses $\msigmarad$, $\msigmaphi$, $\msigmatht$ and $\msigmamean$,
    and biomechanical power $\powerint$.
    Further, biomechanical power due to LV deformation, $\powerint$,
    and hydrodynamic power, $\powerext$, derived from PV data were also compared
    to assess differences during isovolumetric phases.

%

  \subsection{Numerical Solution}
    Discretization of all PDEs and the solution of the arising systems of equations
    relied upon the Cardiac Arrhythmia Research Package framework \citep{vigmond03:_carp}.
    Details on FE discretization \cite{rocha11:_hybrid} as well as
    numerical solution of electrophysiology \cite{vigmond08:_solvers,neic12:_accelerating,neic17:_efficient}
    and electro-mechanics \cite{augustin16:_anatomically} equations have been described in detail previously.
    Both electrophysiolgy and mechanics FE solvers were validated previously
    in N-version benchmark studies \cite{niederer11:_nversion_ep,land15:_nversion}.

  \subsectionend{Evaluation of clinical wall stress assessment}


\section{Results}
  \label{sec:results}


  \subsection{Verification of FE model}
  \label{sec:_verify_fe_results}

\begin{figure}[!ht]
      \centering
      \subfloat[\;(A)]{ \includegraphics[width=0.35\textwidth, keepaspectratio]{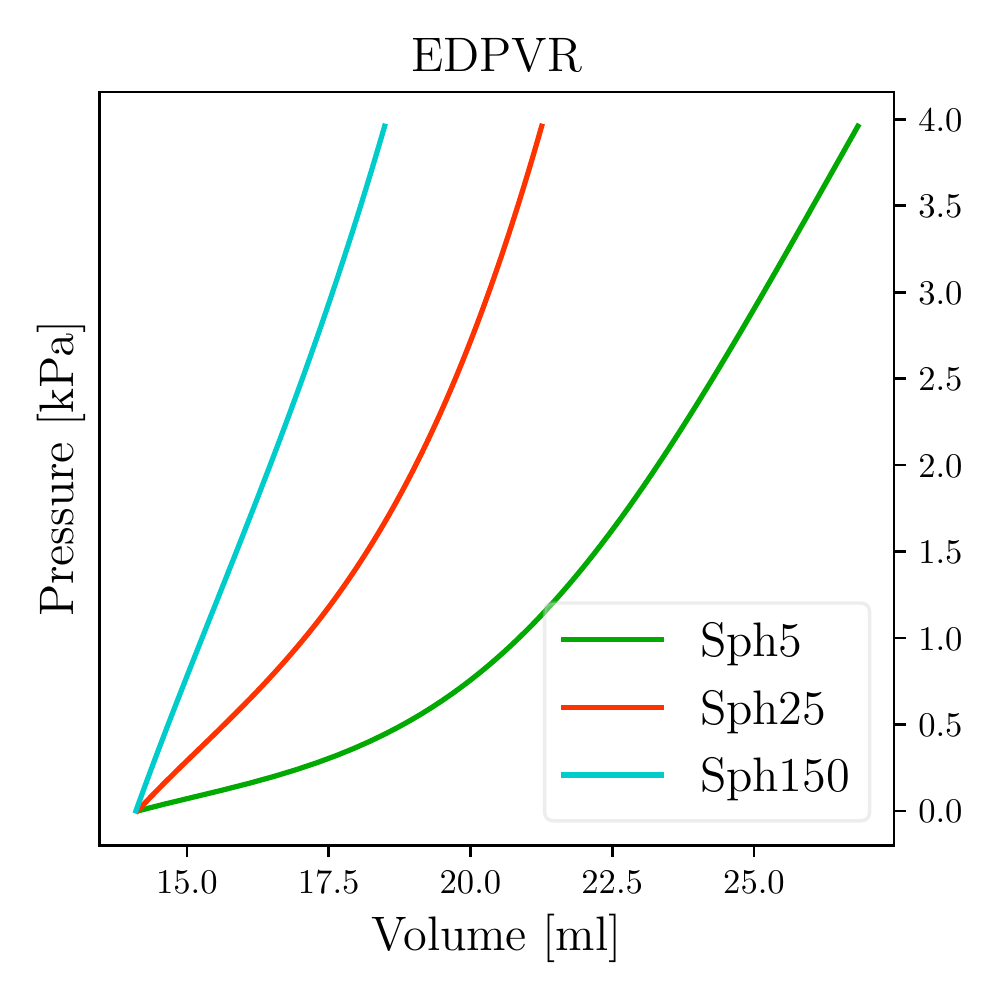}
      	\label{fig:_edpvr_sph}}
      \subfloat[\; (B)]{ \includegraphics[width=0.35\textwidth, keepaspectratio]{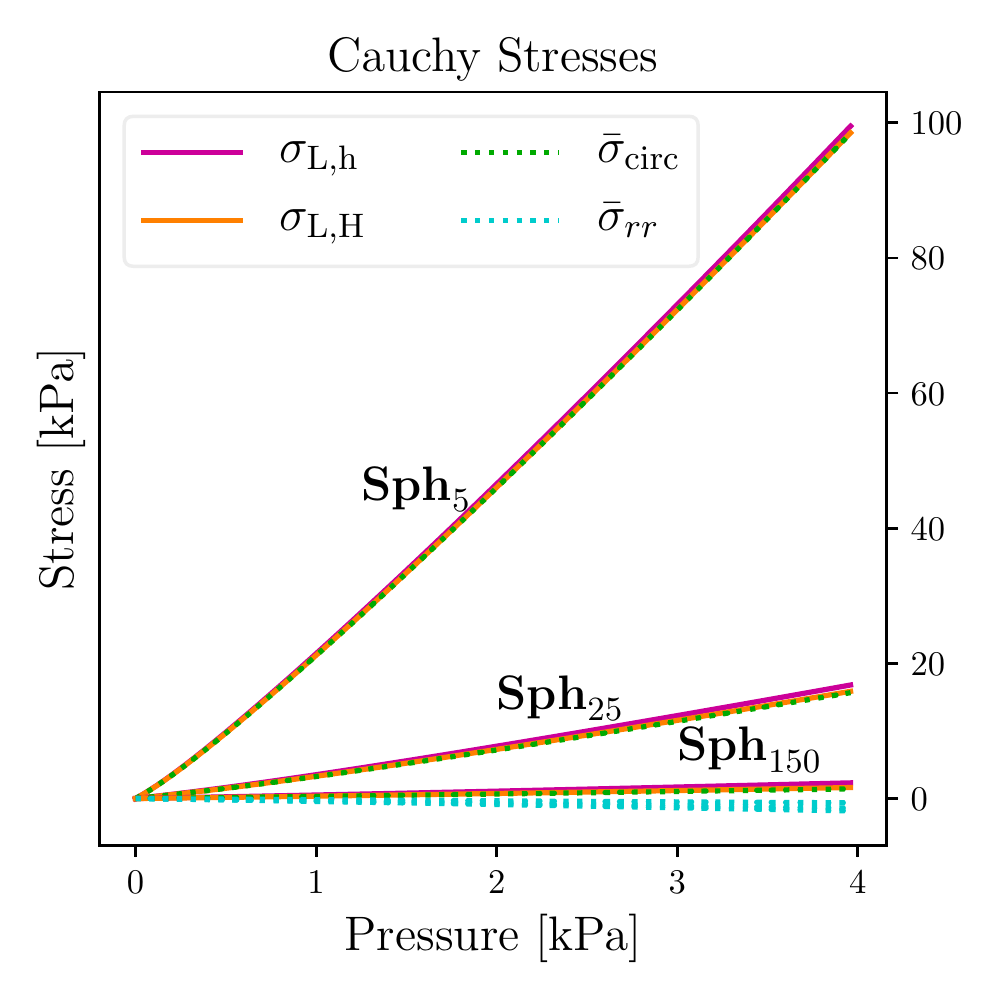} \label{fig:_edpvr_stress}}
      \caption{
     	(A)~EDPVRs for isotropic spherical shell models of varying wall thickness $h$.
      	(B)~Comparison of FE-based mean circumferential and radial stresses, $\msigmacirc$ and $\msigmarad$,
      	with Laplace-based estimations $\sigmaLsmp$ and $\sigmaLext$.}
      \label{fig:_fe_verification}
    \end{figure}

    The FE implementation was verified by performing passive inflation experiments
    with spherical shell models for which Laplace laws are known to be almost exact ($\sphh$)
    or, at least, sufficiently accurate ($\sphH$ and $\sphHH$).
    The resulting EDPVRs and principal components of the Cauchy stress $\sigmatotsph$,
    evaluated in spherical coordinates and globally averaged
    to yield mean stresses $\msigmacirc$ and $\msigmarad$,
    are shown in Figs.~\subref*{fig:_edpvr_sph} and \subref*{fig:_edpvr_stress}.
    A numerical comparison of stresses and work at the maximum pressure of $p=4$~kPa
    is provided in Tab.~\ref{tab:exp:stress} and Tab.~\ref{supp:tab:exp:work} in the supplement.

    Agreement of FE-based mean circumferential stress $\msigmacirc$ with Laplace laws
    was very close,
    that is, $\msigmacirc = \frac{1}{2} (\msigmatht + \msigmaphi) \approx \msigmatht \approx \msigmaphi \approx \sigmaLext \approx \sigmaLsmp$ held.
    With increasing $h$ $\sigmaLext$ provided estimates
    that were closer to the FE-based stress $\msigmacirc$ than $\sigmaLsmp$
    (Tab.~\ref{tab:exp:stress}).
    The simple Laplace overestimated $\msigmacirc$ in $\sphh$, $\sphH$ and $\sphHH$
    by $1.02\%$, $8.32\%$ and $63.19\%$
    whereas with $\sigmaLext$ deviations were much smaller
    with $0.02\%$, $2.3\%$ and $14.58\%$.
    Radial stresses were negligible in the thinner-walled models $\sphh$ and $\sphH$,
    that is, $\msigmarad \ll \msigmacirc$,
    but not in the thick-walled model $\sphHH$
    where $\msigmarad$ amounted to $\approx 43.75\%$ of $\msigmacirc$.
    A comparison of FE-based work $\workint$ to Laplace-based $\workintsmp$ and $\workintext$
    is given in Tab.~\ref{supp:tab:exp:work} in the supplement.

    \begin{center}
      \begin{table}[h!]
        \centering
    \resizebox{\linewidth}{!}{
        \begin{tabular*}{500pt}{@{\extracolsep\fill}l|ccc|ccc|cc@{\extracolsep\fill}}
          \toprule
          Setup & $\msigmarad$ [kPa] & $\msigmatht$ [kPa] & $\msigmaphi$ [kPa] & $\sigmaLsmp$ [kPa] & $\sigmaLext$ [kPa] & $\sigmaLvol$ [kPa]& \#elements & $\bar{\dx}$ [mm] \\
          \midrule
          $\sphh$ & $-1.77$ & $99.46$ & $99.47$ & $100.47$ & $99.48$ &&  $83825$ & $0.65$ \\
          $\sphH$ & $-1.16$ & $15.63$ & $15.63$ & $16.93$ & $15.99$ &&  $40974$ & $1.23$ \\
           $\sphHH$ & $-0.63$ & $1.44$ & $1.44$ & $2.35$ & $1.65$ && $213742$ & $1.70$ \\
          \midrule
          $\lvagu$ & $-0.90$ & $7.05$ & $3.33$ & $5.24$ & $4.40$ & $6.83$ & $420704$ & $1.52$ \\
          $\lvbgu$ & $-0.85$ & $8.71$ & $4.29$ & $5.94$ & $5.08$ & $8.19$ & $332221$ & $1.74$ \\
          $\lvcgu$ & $-0.67$ & $4.47$ & $2.12$ & $3.54$ & $2.76$ & $4.44$ & $456553$ & $1.84$ \\
          $\lvdgu$ & $-0.83$ & $6.97$ & $3.93$ & $5.73$ & $4.88$ & $7.07$ & $394808$ & $1.86$ \\
          \bottomrule
      \end{tabular*}}
        \caption{Comparison of FE-based mean wall stresses $\msigmarad$, $\msigmatht$ and $\msigmaphi$
                in radial, azimuthal and meridional direction, respectively,
              with the Laplace-based wall stress estimates $\sigmaLsmp$, $\sigmaLext$
              and $\sigmaLvol$.
              All stresses refer to the maximum applied pressure of $p=4~$kPa.}
        \label{tab:exp:stress}
      \end{table}
    \end{center}

  \subsectionend{Verification}

\vspace{-1cm}
\subsection{Evaluation of Laplace-based assessment of wall stresses and power}
\label{subsec:applicationlv}
  After verification with spherical shell models
  FE analysis was applied to a validated \emph{in silico} EM LV model
  to compute stresses, power and work during both diastolic and systolic phases.
  Since all assumptions underlying Laplace laws are violated in LV models
  the FE-based results were considered the ground truth and, thus,
  could be used to gauge the accuracy of Laplace-based assessment of LV mechanics.

\subsubsection{Passive inflation of LV models}

\begin{figure}[!ht]
     \centering
     \includegraphics[width=1.0\textwidth, keepaspectratio]{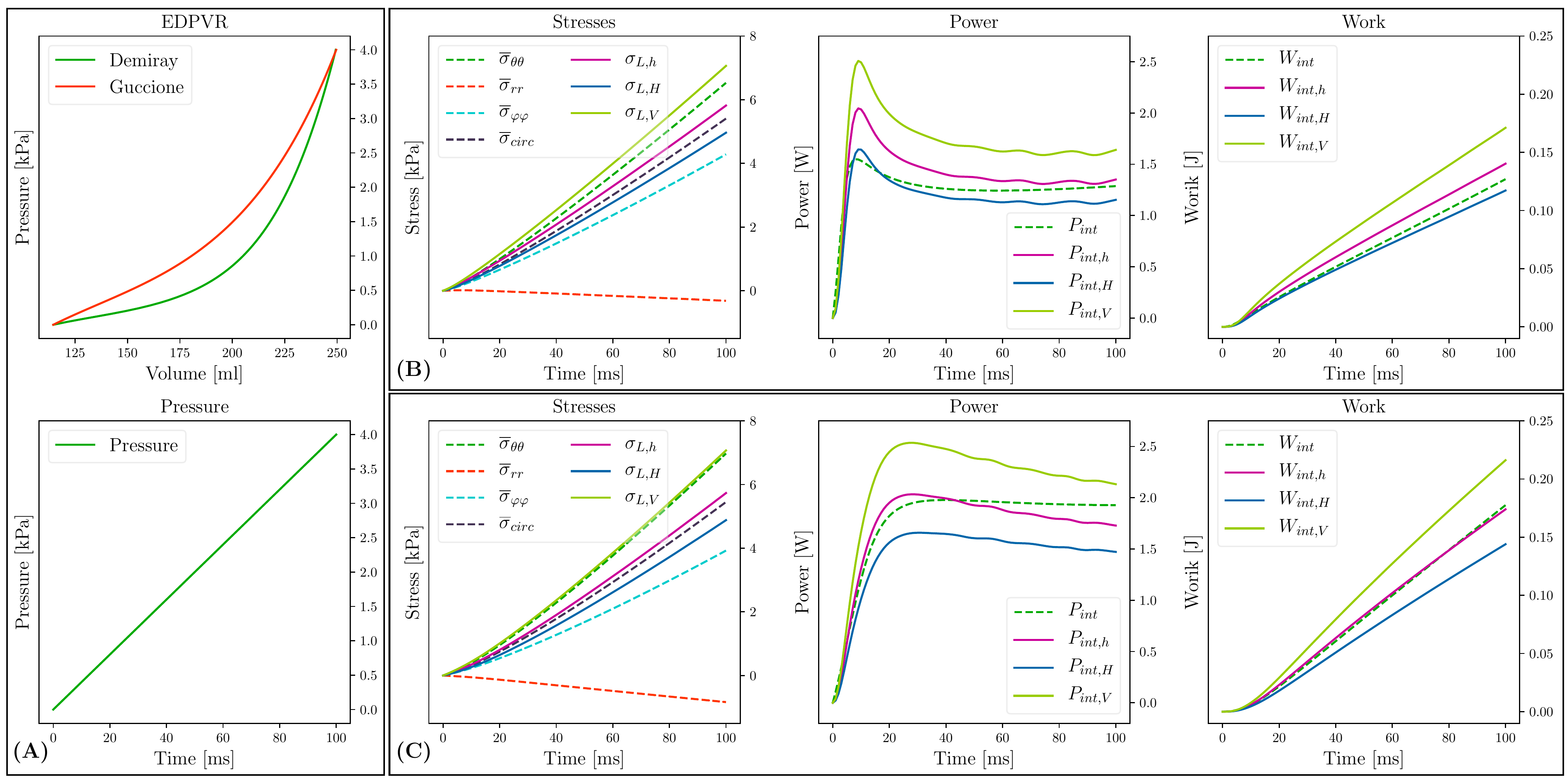}
     \caption{(A) Loading protocol. (B) Stresses, power and work for anisotropic model.
     Data are shown for model $\lvd$.}
     \label{fig:lv_to_laplace}
   \end{figure}

   LV models $\lva$--$\lvd$ were inflated following the same protocol as in Sec.~\ref{sec:verificationoffemodel}
   (see Fig.~\ref{fig:lv_to_laplace}.A). The temporal evolution of FE- and Laplace-based stresses, power and work
   are shown in Fig.~\ref{fig:lv_to_laplace} for model $\lvd$.
   Minor quantitative differences to other models $\lva$--$\lvc$ were observed,
   but qualitatively the overall behavior was identical.
   Stresses at $p=4$~kPa are summarized in Tab.~\ref{tab:exp:stress},
   incurred work is given in Tab.~\ref{supp:tab:exp:work} in the supplement.

\subsubsection{Analysis of LV cycle experiments}
\label{sec:_LV_cycle_analysis}

\begin{figure}[!ht]
  \centering
  \includegraphics[width=1.0\textwidth, keepaspectratio]{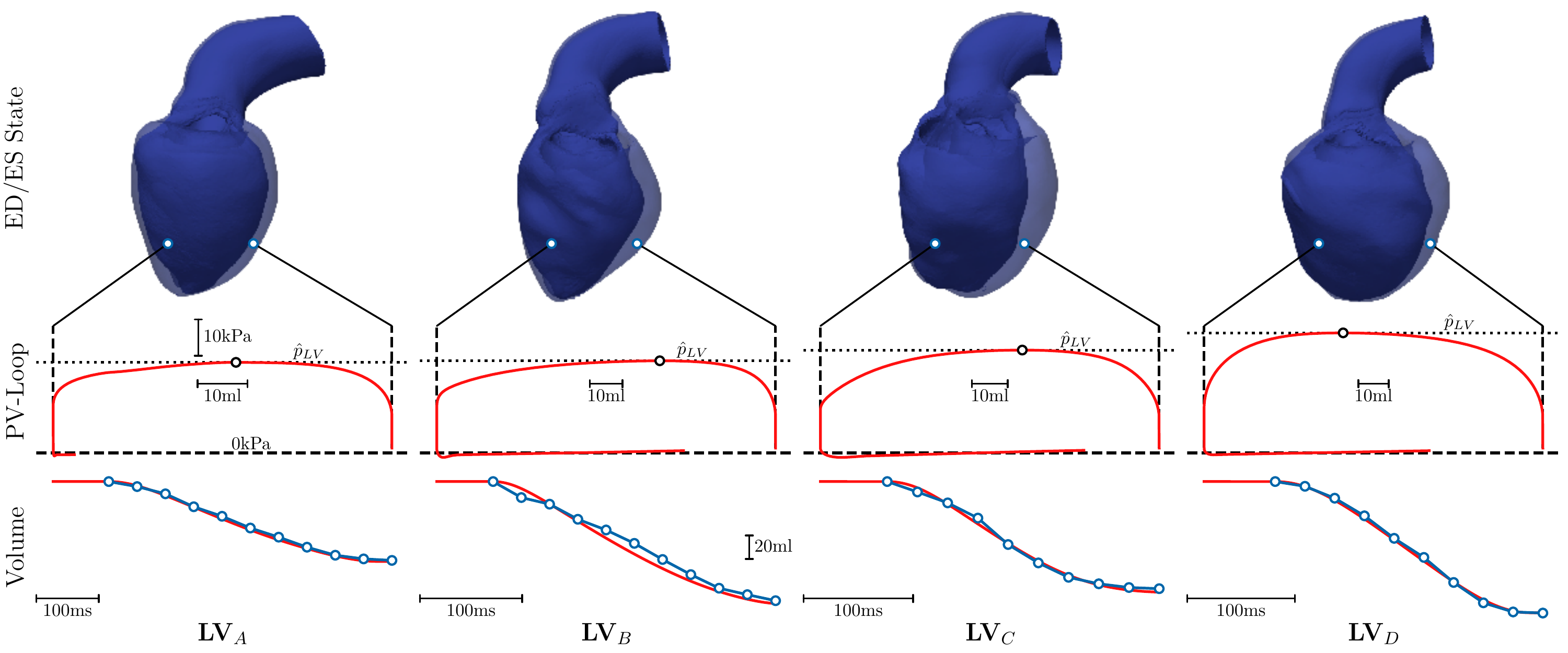}
  \caption{Fitting of EM LV models (red traces) using Cine-MRI-based volume data
  	(blue traces)   and estimated LV peak pressures, $\plvpeak$, as input.
    Top panels show LV anatomy in end-diastolic (transparent blue) and end-systolic (solid blue) configuration.
    }
  \label{fig:_working_lv_validation}
\end{figure}

Using Cine-MRI-based LV volume traces and estimated $\plvpeak$ as inputs
the models $\lva$--$\lvd$ were fitted over the cycle phases isovolumetric contraction (IVC), ejection
and isovolumetric relaxation (IVR) (Fig.~\ref{fig:_working_lv_validation}).
All models replicated the clinical metrics of interest
such as SV, EF or peak aortic pressure $\paopeak$ with sufficient accuracy ($<5\%$).

\begin{figure}[!ht]
  \centering
  \includegraphics[width=1.0\textwidth, keepaspectratio]{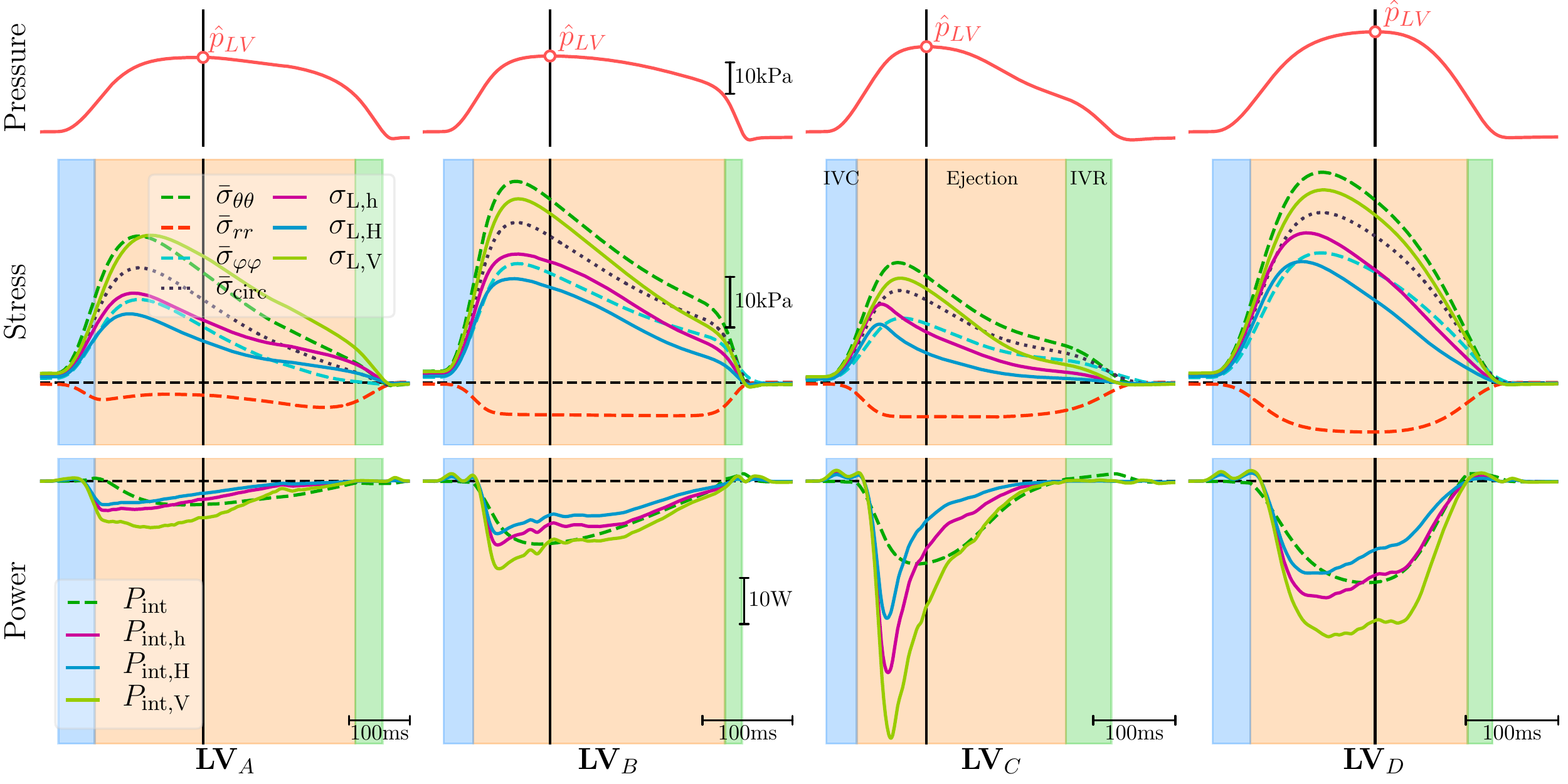}
  \caption{Comparison of FE-based computation of stresses ($\msigmaphi$, $\msigmatht$, $\msigmacirc$ and $\msigmarad$) and power $\powerint$
    with Laplace-based estimates of stress, $\sigmaLsmp$, $\sigmaLext$ and $\sigmaLvol$, and
    power $\powerintsmp$, $\powerintext$ and $\powerintvol$.
    Top panels show the time course of pressure $p$ in the LV endocardium.
    The solid black vertical line indicates the instant, $\tplvpeak$, when peak pressure in the LV, $\plvpeak$, occurs.}
  \label{fig:_working_lv_power}
\end{figure}
\begin{figure}[!ht]
  \centering
  \includegraphics[width=1.0\textwidth, keepaspectratio]{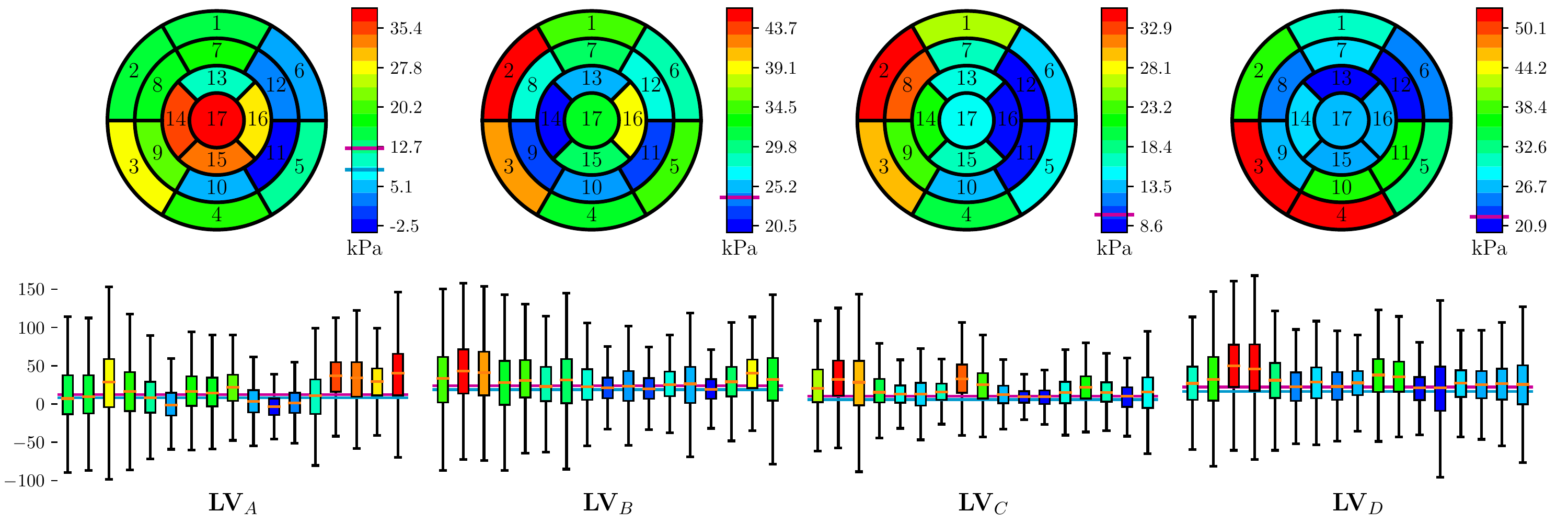}
  \caption{Statistical analysis of the circumferential stress $\sigmacirc$ at the instant of peak pressure.
    The top row shows $\sigmacirc$ averaged over the corresponding LV segment and the
    bottom row shows the variation of $\sigmacirc$ in each LV segment.}
  \label{fig:_lv_stress_analysis}
\end{figure}

Fig.~\ref{fig:_working_lv_power} compares
the time course of the averaged FE-based quantities
azimuthal, meridionial, radial and circumferential mean stresses,
$\msigmaphi$, $\msigmatht$, $\msigmarad$
and $\msigmacirc = \frac{1}{2}(\msigmaphi+\msigmatht)$,
 respectively
and power $\powerint$ to the Laplace-based estimation
of stresses $\sigmaLstar$ and power $\powerintstar$.
In all cases, the Laplace-based stresses $\sigmaLsmp$ and $\sigmaLext$
tended to underestimate the FE-based mean circumferential stress $\msigmacirc$,
being closer to the azimuthal stress $\msigmaphi$,
whereas $\sigmaLvol$ overestimated $\msigmacirc$ and was closer to $\msigmatht$.
Further, both Laplace stresses or globally averaged mean stresses deviate noticeably
from the true local stresses acting at a given location (Fig.~ \ref{fig:_lv_stress_analysis}).

Laplace-based power estimates $\powerintsmp$, $\powerintext$ and $\powerintvol$
were qualitatively comparable to the exact FE-based $\powerint$,
but quantitatively marked discrepancies were observed.
The time course of Laplace-based power showed both a faster onset and decay
with an early peak in power.
Quantitative differences between the Laplace estimates were also significant
with $\powerintvol > \powerintsmp > \powerintext$.
Around the instant $\tplvpeak$ deviations were
in the range of 
$-2.41/+2.92$, $-6.22/-0.58$, $-9.19/+9.34$ and $-7.10/+8.28$~W
for $\lva$, $\lvb$, $\lvc$ and $\lvd$, respectively.
The relative IHP marker led to large deviations (see Fig.~\ref{supp:fig:_working_lv_power:work_only} in the supplement)
from the true FE-based mechanical power,
even around the instant of $\plvpeak$ the IHP marker was intended for.

A numerical comparison between markers of LV peak power is given in Tab.~\ref{tab:peakpower}.
Differences between true mechanical peak power $\powerintpeak$
and $\powerint$ at the instant of peak pressure, $\tplvpeak$, were minor,
with the maximum difference being $\abs{\powerintpeak - \powerint(\tplvpeak)}< 0.19$~W or $2.15\%$.
Laplace estimation of $\powerint$, evaluated at $\tplvpeak$,
misestimated $\powerintpeak$ by $5.1\%$ - $54.7\%$.
Interestingly, $\powerintsmp$ performed better than $\powerintext$ in all cases,
but with non-negligible maximum relative errors of
$23.37\%$, $32.23\%$, $18.51\%$ and $7.68\%$ for $\lva$, $\lvb$, $\lvc$ and $\lvd$, respectively.
$\ihp$ overestimated $\powerintpeak$ significantly,
in the range between $24.10\%$ to $140.28\%$.
Since the mechanical power developed during isovolumetric phases was marginal
(see Fig.~\ref{supp:fig:int_vs_ext_power} in the supplement),
the most accurate estimate of $\powerint$ is obtained
by computing $\powerext$ from standard hemodynamic PV data.
In terms of peak mechanical power the difference $\abs{\powerintpeak - \powerextpeak}$ was less than $0.5 \;W$ or $7.18 \%$ in all cases
where $\powerextpeak$ can be estimated with high accuracy
by taking the product of peak power and flow, $\plvpeak \cdot \qpeak$.

\begin{table}[ht]
	\centering
	\resizebox{\linewidth}{!}{
	\begin{tabular}{l|ccccc|ccc|cc|cc}
		\toprule
		& $\powerintpeak$ & $\powerint(\tplvpeak)$ & $\powerextpeak$ & $\powerext(\tplvpeak)$ & $\plvpeak \cdot \qpeak$
		& $\powerintsmp(\tplvpeak)$ & $\powerintext(\tplvpeak)$ & $\powerintvol(\tplvpeak)$ & $\ihp$ & $\ehp$ & $\powereff$ & $\powereffclin$ \\
		& [W] & [W] & [W] & [W] & [W] & [W] & [W] & [W] & [W] & [W] & [1] & [1]\\
		\midrule
		$\lva$ & $-5.13$ & $-5.02$   & $-5.50$  & $-5.42$  & $-5.54$  & $-3.93$  & $-2.62$  & $-7.94$  & $-6.50$  & $-2.56$ & $0.49$ & $0.39$  \\
		$\lvb$ & $-13.64$ & $-13.52$ & $-13.56$ & $-13.48$ & $-13.59$ & $-9.24$  & $-7.30$  & $-12.94$ & $-32.76$ & $-6.00$ & $0.43$ & $0.18$ \\
		$\lvc$ & $-17.94$ & $-17.75$ & $-17.44$ & $-17.39$ & $-17.48$ & $-14.62$ & $-8.56$  & $-27.09$ & $-22.27$ & $-6.30$ & $0.35$ & $0.28$ \\
		$\lvd$ & $-22.01$ & $-21.92$ & $-21.80$ & $-21.74$ & $-21.84$ & $-20.32$ & $-14.82$ & $-30.20$ & $-29.59$ & $-8.02$ & $0.36$ & $0.27$ \\
		\bottomrule
	\end{tabular}
	}
	\caption{Comparison of FE-based peak mechanical power $\powerintpeak$
		with different estimates that were all evaluated at the instant of peak pressure, $\tplvpeak$:
		$\powerint(\tplvpeak)$, $\powerext(\tplvpeak)$, $\powerintsmp(\tplvpeak)$, $\powerintext(\tplvpeak)$, $\powerintvol(\tplvpeak)$;
		peak hydrodynamic power $\powerextpeak$; product of peak pressure and peak flow in the LV $\plvpeak \cdot \qpeak$;
		relative internal heart power marker $\ihp$ and external heart power $\ehp$;
		cardiac power efficiency $\powereff$ and its $\ihp$-based approximation  $\powereffclin$.}
	\label{tab:peakpower}
\end{table}


\section{Discussion}
  \label{sec:discussion}

Wall stress and mechanical power generated by the LV are considered important biomarkers
that promise potential clinical utility for diagnosis and as a predictor of post-treatment LV remodeling
after interventions 
\cite{alter16:_wallstress,fernandes17:_beyond}.
Moreover, the modelling of stresses and power would allow to gain an improved understanding of mechanisms
that contribute to adverse remodelling. Laplace analysis would have the charme that inputs such as
$p$, $r$, $h$, $\volmyo$ and $\volcav$ are accessible within routine clinical procedures.
However, Laplace analysis is based on a global force balance calculation
and relies upon simplifying assumptions on LV shape, tissue structure and biomechanical behavior.
This study attempts to establish validity, accuracy and potential limitations of Laplace analysis
of stresses and mechanical power generated by the LV by comparing against a FE model
for which these quantities can be determined with high accuracy.

\subsection{FE verification}

FE computation of stresses and mechanical power was verified
by performing passive inflation experiments
with geometrically well defined spherical shell models of varying wall width
for which Laplace laws hold with sufficient accuracy.
FE computed circumferential stresses $\sigmacirc$ in all models
agreed closely with the Laplace stresses $\sigmaLstar$
(see Fig.~\ref{fig:_fe_verification} and Tab.~\ref{tab:exp:stress}).
As expected, with increasing $h$, deviations became more pronounced
and the thick-walled Laplace stresses $\sigmaLext$ agreed closer with FE stresses than the standard Laplace stress  $\sigmaLsmp$.
In terms of work expended,
more noticeable discrepancies were observed between $\workint$ and Laplace-based $\workintstar$ (see Tab.~\ref{supp:tab:exp:work} in the supplement).
However, since the agreement between FE-computed internal work $\workint$
and external work $\workext$ was essentially perfect,
as expected on grounds of conservation of energy,
we concluded that our FE implementation
for evaluating stresses, power and work is correct
and that the observed deviations are rather attributable to inherent inaccuracies
in the Laplace approximations.
In particular, we consider the mean strain rate approximation in Eq.~\eqref{equ:lol:ipow}
and the omission of radial stresses likely candidate causes.

\subsection{Laplace versus FE-based stress and power analysis}

The validated high resolution \emph{in silico} model served as a reference
for evaluating the accuracy of the Laplace-based approximation of
$\sigmatot$, $\powerint$ and $\workint$.
While the FE models which were built from, fitted to and validated against clinical data,
may deviate from clinical data within the limits of clinical data uncertainty,
for assessing Laplace analysis, the FE model represents the ground truth
as it provides accurate data on stresses $\sigmatot(\vx,t)$ and strains $\strain(\vx,t)$
which can serve to compute local power and work densities $\powerintdensity(\vx,t)$ and $\workintdensity(\vx,t)$, respectively,
as well as global $\powerint$, $\workint$ and $\workext$  with highest possible accuracy.
All input parameters needed for Laplace analysis can be derived from the FE model
with higher accuracy than what is achievable clinically.
In this regard, the application of Laplace analysis to the \emph{in silico} model
can be considered a best case scenario.

\subsubsection{Wall stress in the LV}

Wall stress $\sigmatot(\vx)$ in the LV is a tensorial quantity
that varies in space (see Fig.~\ref{fig:_lv_stress_analysis}).
The tensor comprises six independent components
whereas Laplace stresses $\sigmaLstar$ provide only one scalar stress value
representing a global circumferential or hoop stress, $\sigmacirc$.
While $\sigmacirc$ is equivalent to $\sigmaphi$ and $\sigmatht$ in a thin-walled spherical shell such as $\sphh$ (see Tab.~\ref{tab:exp:stress})
this is not the case in the LV
as there is no direct equivalence to any component of $\sigmatot$.
As shown in a FE modeling study by Zhang~et~al.~\cite{zhang2011comparison},
the correlation of Laplace stresses to fiber and cross-fiber stresses is poor.
Conceptually, the force balance consideration used in the derivation of Laplace laws suggests
that Laplace stresses are most likely representative of the mean stresses
in the longitudinal-circumferential plane,
$\msigmacirc = \frac{1}{2}\left( \msigmaphi + \msigmatht \right)$.
Indeed, a fair qualitative agreement was observed between $\msigmacirc$ and  $\sigmaLstar$ during passive LV inflation
as illustrated in Fig.~\ref{fig:lv_to_laplace}.
During ejection the time course of $\sigmaLstar(t)$ followed a similar trend as $\msigmacirc$
although waveforms deviated to different degrees owing to the marked differences in the LV anatomies.
However, quantitatively discrepancies were significant
during both passive inflation and over a LV cycle
as evident in Fig.~\ref{fig:lv_to_laplace}.B and in the stress panels of Fig.~\ref{fig:_working_lv_power}
with substantial differences in stress magnitudes between the various Laplace laws
and the global circumferential mean stress
with  $\sigmaLvol > \msigmacirc > \sigmaLsmp > \sigmaLext$.

Besides the fundamental problem of stress heterogeneity and tensorial properties of LV wall stress,
Laplace calculations are afflicted with significant uncertainties.
The meaning of geometric parameters $r$ and $h$
required for the evaluation of Eqs.~\eqref{equ:lol:ext} or \eqref{equ:lol:smp} is ambiguous
when applied to the LV
which deviates in shape markedly from a spherical shell.
Therefore, $r$ and $h$ must be determined from averaging over a number of short axis Cine MRI scans
to find representative values.
Due to longitudinal shortening additional averaging occurs
as different slices of the heart are being imaged during ejection.
Thus, the determination of parameters $r$ and $h$ cannot be unique
as the particular method employed for averaging,
such as the one described in Eq.~\eqref{equ:_mean_r_h},
influences, to some extent, the results.
Using  Eq.~\eqref{equ:lol:vol} seems to circumvent this problem
since $\volcav$ and $\volmyo$ are used as inputs
which may be determined uniquely for the LV.
However, in our simulations $\sigmaLvol$ led to larger misestimations than $\sigmaLsmp$ and $\sigmaLext$.

It is well known that Laplace-based calculation of stresses is afflicted
with various inaccuracies \cite{moriarty1980law}.
Nonetheless, Laplace-based calculation of LV wall stresses has been used in clinical studies as a diagnostic criterion \cite{alter16:_wallstress}.
However, according to observations in this study
based on an \emph{in silico} model
and in line with other studies \cite{zhang2011comparison},
the scope for clinical applications appears narrow.
Laplace stresses may provide information of diagnostic value,
but, if so, rather as an empirical than a mechanistic marker.
As a biomarker representing LV wall stresses in a physical sense
Laplace-based calculations suffer from severe fundamental limitations.

\subsubsection{Mechanical heart power and power efficiency}

Mechanical heart power  $\powerint$ and cardiac power efficiency $\powereff$ defined as
the ratio between peak mechanical power expended by the LV, $\powerintpeak$,
and the mean hydrodynamic power delivered to the arterial system, $\ehp$,  have been proposed recently
as a diagnostic marker \cite{fernandes17:_beyond}.
On grounds of conservation of energy the global mechanical power $\powerint$ expended by the LV
and the hydrodynamic power transferred to the LV blood pool, $\powerext$, must be equal.
Discrepancies may occur due to isovolumetric phases
during which hydrodynamic power is close to zero,
but mechanical power is expended by the LV to some extent
as conformational changes of the LV myocardium
and the shape of the LV cavity occur.
However, in all models studied $\powerint$
during isovolumetric phases was negligible (see Fig.~\ref{supp:fig:int_vs_ext_power} in the supplement).
This does not conflict with experimental studies
providing evidence of heterogeneous circumferential strains,
longitudinal shortening and wall thickening during IVC \cite{ashikaga09:_transmural_ivc}.
Qualitatively similar behavior is observed in our \emph{in silico} models,
but magnitude and velocity of strain development is much smaller
during IVC than during ejection.
Thus, the strain rate tensors $\straindot$ remained small during IVC
and  mechanical power expenditure was minor.
In all LV models under study
global mechanical power  $\powerint$ and the hydrodynamic power in the LV cavity $\powerext$
were virtually identical (see Tab.~\ref{tab:peakpower} and Fig.~\ref{supp:fig:int_vs_ext_power} in the supplement).
Hence, mechanical heart power can be determined either
by analyzing the deformation of the LV myocardium
or from PV relations in the LV.

The estimation of $\powerint$ is feasible
directly from LV deformation either by using FE models or,
as suggested in \cite{fernandes17:_beyond}, based on Laplace's law
where the latter approach is more readily applicable in the clinic.
However, when global LV power is of interest,
Laplace-based approaches do not seem to offer any additional benefits
over more standard approaches
relying on hemodynamic data for a number of reasons.

First of all, the evaluation of $\powerintstar$ based on Eq.~\eqref{equ:lol:ipow}
introduces a systematic error which leads to a misestimation of the actual $\powerint$,
even in the spherical shell models,
since any work expended in the radial direction is ignored. 
Laplace's law takes into account only circumferential stresses
and neglects any radial stresses.
As shown in Fig.~\ref{fig:_fe_verification},
this simplification is only well justified in thin-walled structures such as $\sphh$,
but introduces pronounced discrepancies for increased $h$
(see passive inflation experiments in Tab.~\ref{tab:exp:stress} and Tab.~\ref{supp:tab:exp:stress}
in the supplement as well as $\msigmarad$ traces in Fig.~\ref{fig:_working_lv_power}).
Secondly, in addition to the parameters needed for wall stress estimation
which are afflicted with substantial uncertainties as discussed above,
the parameters $V_{\mathrm{myo}}$ and $\straindot$ are required.
Using the approximation given by Eq.~\eqref{supp:equ:app:strainrate} in the supplement
the estimation of $\straindot_{\rm{circ}}$
requires that both inner and outer radii $r$ and $R$ of the LV can be tracked
with sufficient temporal resolution and accuracy.
However, as evidenced in Fig.~\ref{fig:_working_lv_power},
even when evaluated in an \emph{in silico} model where tracking of these quantities is feasible with the highest possible accuracy,
the overall accuracy of the method is rather poor
with significant under- or overestimation of the true $\powerint$,
depending on whether $\powerintsmp$, $\powerintext$ or $\powerintvol$ is used
and whether an early or late phase of ejection is considered (see Fig.~\ref{fig:_working_lv_power}).

The evaluation of cardiac power efficiency $\powereff$ or $\powereffclin$
requires only point estimates of peak mechanical power.

Following \cite{fernandes17:_beyond},
this is feasible by assuming that $\powerintpeak$ occurs at the instant of peak pressure, $\tplvpeak$.
Consistent with expectations based on Laplace's law
this was not the case in any of our LV models.
As $p \propto \sigma / \left( r/h \right)$
peak pressure $\plvpeak$ and peak stress
would only coincide under isometric conditions.
In the contracting LV during ejection, the ratio $\left( r/h \right)$ decreases,
thus facilitating a further increase of $p$ beyond the instant of peak pressure
(see pressure and stress panels in Fig.~\ref{fig:_working_lv_power}).
Nonetheless, the instants of peak power and peak pressure fell sufficiently close together
with $\abs{\tplvpeak - \tpowerpeak}$ of $32$, $12$, $11$ and $9$ ms
for $\lva$, $\lvb$, $\lvc$ and $\lvd$, respectively.
Indeed, inspection of Tab.~\ref{tab:peakpower} and the power panels in Fig.~\ref{fig:_working_lv_power} suggests
that the Laplace-based estimation of $\powerintpeak$ seems feasible
by evaluating power at the instant of peak pressure
(compare $\powerintpeak$, $\powerint(\tplvpeak)$,
$\powerintsmp(\tplvpeak)$,$\powerintext(\tplvpeak)$ and $\powerintvol(\tplvpeak)$
in Tab.~\ref{tab:peakpower}),
albeit with inferior accuracy compared to estimations based on hemodynamic PV data.

Alternatively,
the simpler $\powereffclin$ marker can be used as in \cite{fernandes17:_beyond}
which relies on $\ihp$ and does not require an estimation of $\straindot$.
While simpler, its use brings about a number of drawbacks.
Since $\straindot$ is ignored $\ihp$ is only a relative marker
that is non-linearly related to $\powerint$.
Therefore, $\ihp$ provided highly inconsistent relative estimates of $\powerint$
with errors varying in the range from $24.1\%$ to $140.28\%$ (see Tab.~\ref{tab:peakpower}).
Thus, $\ihp$ as an indicator of $\powerint$ appears to be of insufficient accuracy
even for clinical applications of modest accuracy demands.
Overall, the scope for Laplace-based power estimation as proposed in  \cite{fernandes17:_beyond}
seems limited
as standard methods based on hemodynamic data
are afflicted with less uncertainty, offer higher accuracy and
are easier to evaluate.
As shown in Tab.~\ref{tab:peakpower}, $\powerintpeak$ is straight forwardly approximated
-- with higher accuracy than any Laplace-based method --
as the product of peak pressure and flow, $\hat{p} \cdot \hat{q}$.

The mechanical power generated by the LV is an indicator of metabolic demands.
Local wall stresses and power densities governing energetic demand and supply ratios in the LV myocardium
are known to play important roles as drivers of remodeling in the pressure-overloaded LV of AS patients.
However, analogous to the stresses shown in Fig.~\ref{fig:_working_lv_power}
the distribution of power density $\powerintdensity(\vx,t)$ in the LV wall
is highly heterogeneous as well with significant regional variability around the global mean power density.
In this view Laplace-based global markers derived from mechanical deformation such as $\msigmacirc$ or $\powerintstar$
are not representative of local stresses and power within the LV myocardium
and appear to offer limited insight and predictive power beyond standard PV analysis.

An accurate representation of local mechanical stresses $\sigmatot(\vx,t)$ and power $\powerintdensity(\vx,t)$
over a cardiac cycle depends on reliable sets of strains $\strain(\vx,t)$.
While techniques for measuring strains in 3D throughout the LV myocardium
are available \cite{ibrahim11:_tagging},
such recordings are not part of clinical routine,
their analysis requires expensive non-trivial post-processing
and spatio-temporal resolution and accuracy are limited.
A carefully fitted and validated FE-based EM LV model
which replicates a patients physiology in terms of PV relations as well as LV kinematics,
provides accurate data on strains $\strain(\vx,t)$ at a high spatio-temporal resolution.
Using an appropriate parameterized patient-specific constitutive model
such as given in Eq.~\eqref{eq:guccioneStrainEnergy},
$\strain(\vx,t)$ can be used to compute LV wall stresses $\sigmatot(\vx,t)$
and $\powerint$ or any other stress-related biomarker efficiently with high accuracy.
Such models are able to provide either global power $\powerint(t)$,
but also fine-grained distributed power density $\powerintdensity(\vx,t)$.
A spatio-temporal view on
$\strain(\vx,t)$, $\sigmatot(\vx,t)$ and $\powerintdensity(\vx,t)$ in the LV
may provide additional insights
as regions of elevated strain, stress or power
are assumed to be implicated in the mechanisms driving remodeling
in the pressure overloaded LV \cite{grossman75:_wall,sadoshima97:_cellular}.

\section{Conclusions}

Laplace estimates of LV wall stress are able to provide a rough approximation
of global mean stress in the circumferential-longitudinal plane of the LV.
However, according to FE results
spatial heterogeneity of stresses in the LV wall is significant,
leading to major discrepancies between local stresses and global mean stress.
Assessment of mechanical power with Laplace methods is feasible,
but these are inferior in accuracy compared to FE models and do not offer any benefits
compared to standard methods based on hemodynamic data.
In this view, the scope for Laplace-based analysis in clinical applications seems narrow.
The accurate assessment of stress and power density distribution in the LV wall
is only feasible based on patient-specific FE modeling.

%
\bibliographystyle{plainnat}
\bibliography{wileyNJD-AMA}
\clearpage


\section{Supplementary Material}


\subsection{Derivation of Laplace's law}
\label{supp:subsec:stresses}

\begin{figure}[!ht]
  \centering
  \subfloat[\;Geometrical setup.]{ \includegraphics[scale=0.7]{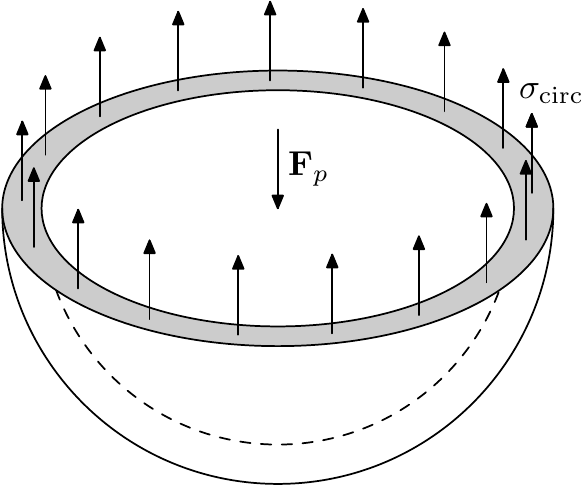} \label{supp:fig:sph:bof} }
  \hspace{1.5cm}
  \subfloat[\;Geometrical setup.]{ \includegraphics[scale=0.7]{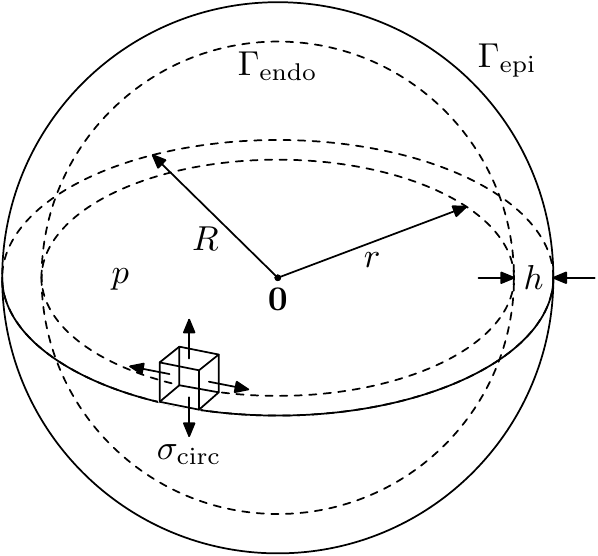} \label{supp:fig:sph:geom} }
  \hspace{1.5cm}
  \subfloat[\;Spherical coordinate system.]{ \includegraphics[scale=0.7]{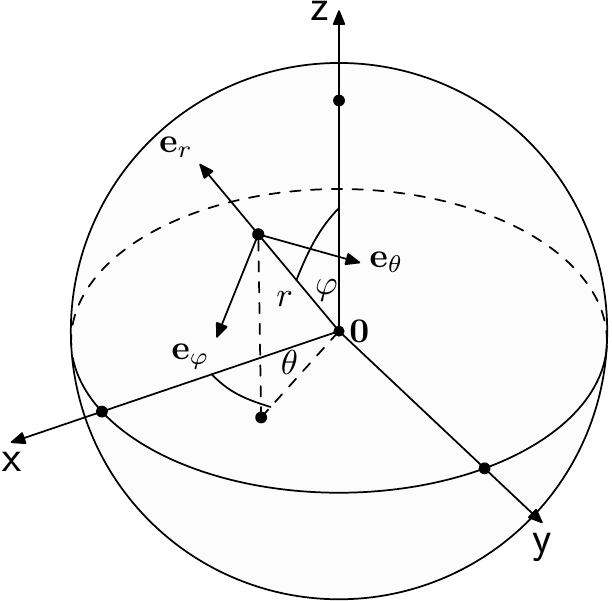} \label{supp:fig:sph:csys} }
  \caption{Balance of forces in thin-walled spherical shell models, spherical coordinate system and displacement boundary conditions.}
  \label{supp:fig:sph}
\end{figure}

To derive the law of Laplace we consider a spherical shell of inner radius,
$r$ and thin walls of a given thickness, $h$. Inflating the shell by applying a pressure,
$p$, within the shell's cavity, induces deformation which causes the buildup of stresses
within the wall. If we consider one half of the sphere, the total force $\vec{F}_p$
acting on the inner surface must be balanced with the total force acting  over the cut
surface (see Fig.~\subref*{supp:fig:sph:bof}). Due to spherical symmetry,
the circumferential stresses $\sigmacirc(\rho)$ at any radius $r \le \rho \le R$
must be the same and the shear stress is zero.
Integrating circumferential stresses over the cut surface yields the total force
balancing the force due to the applied pressure.
That is, we have
\begin{equation}
  \label{supp:equ:app:bof}
  p \, r^2 \, \pi \stackrel{!}{=} (R^2 - r^2) \, \sigmacirc \, \pi.
\end{equation}
Assuming that $h \ll r$, radial stresses are small compared to circumferential stresses,
$\sigmarad \ll \sigmacirc$, and the total stress tensor is approximated by
\begin{equation}
  \label{supp:equ:app:sig:sph:smp}
  \sigmatot
  = \left(
  \begin{array}{ccc}
  \sigma_{xx} & \sigma_{xy} & \sigma_{xz} \\
  \sigma_{yx} & \sigma_{yy} & \sigma_{yz} \\
  \sigma_{zx} & \sigma_{zy} & \sigma_{zz} \\
  \end{array}
  \right)
  = \mathbf{P} \left(
  \begin{array}{ccc}
  \sigma_{rr} & \sigma_{r \varphi} & \sigma_{r \theta} \\
  \sigma_{\varphi r} & \sigma_{\varphi \varphi} & \sigma_{\varphi \theta} \\
  \sigma_{\theta r} & \sigma_{\theta \varphi} & \sigma_{\theta \theta} \\
  \end{array}
  \right) \mathbf{P}^{\top}
  \approx \mathbf{P} \left(
  \begin{array}{ccc}
  0 & 0 & 0 \\
  0 & \sigmacirc & 0 \\
  0 & 0 & \sigmacirc \\
  \end{array}
  \right) \mathbf{P}^{\top}
\end{equation}
with respect to the spherical coordinate system and the projection matrix $\mathbf{P} = (\vec{e}_r, \vec{e}_{\varphi}, \vec{e}_{\theta})^{\top}$,
see Fig.~\subref*{supp:fig:sph:csys}.
Note that $\sigmatot$ in a spherical shell differs from a stress tensor in the LV
in various ways.
Unlike in the LV, stresses in circumferential and meridonial/longitudinal direction are equal
whereas in the LV longitudinal stresses tend to be larger than circumferential stresses.
Further, the assumption $r \ll h$ is not justified, rather $r \approx h$ holds.
Thus radial stresses in the LV are non-negligible, that is, $\sigmarad$ is at an order of magnitude
comparable to $\sigmacirc$.

\subsubsection{Laplace's law for a thick-walled sphere}
\label{supp:subsubsec:extlaplace}
From Eq.~\eqref{supp:equ:app:bof} the circumferential stress in a thick-walled sphere 
is found as
\begin{equation}
  \label{supp:equ:app:lol:ext}
  \sigmacirc = \frac{p \, r^2}{(R^2 - r^2)} = \frac{p \, r^2}{(R - r)(R + r)} = \frac{p \, r^2}{h (2 \, r + h)}
  = \frac{p \, r^2}{2 \, h \, r \left( 1 + \frac{h}{2 \, r} \right) }
  = \frac{p \, r}{2 \, h \left( 1 + \frac{h}{2 \, r} \right) } = \sigmaLext
\end{equation}
which we denote as $\sigmaLext$.


\subsubsection{Laplace's law for a thin-walled sphere}
\label{supp:subsubsec:smplaplace}
Using assumption~\textbf{(A3)}, i.e. $h/r \ll 1$, we have $\left(1+\frac{h}{2 \, r}\right) \approx 1$
which yields the simple law of Laplace for a thin-walled sphere given by
\begin{equation}
  \label{supp:equ:app:lol:smp}
  \sigmacirc = \frac{p \, r}{2 \, h \left( 1 + \frac{h}{2 \, r} \right)}
  \approx \frac{p \, r}{2 \, h} = \sigmaLsmp
\end{equation}
which we denote as $\sigmaLsmp$.

\subsubsection{Volume-based stress}
\label{supp:subsubsec:volstress}
Since radius $r$ and wall thickness $h$ are two quantities which are not always available
or are hard to determine for a general geometry like a LV, we rewrite Eq.~\eqref{supp:equ:app:lol:ext}
in terms of the cavity volume $\volcav$ and the myocardial volume $\volmyo$. For the spherical shell
geometry there holds $\volcav = \frac{4}{3} \pi r^3$ and $\volcav = \frac{4}{3} \pi \left( R^3 - r^3 \right)$ which entails
$r = \left( \frac{3 \, \volcav}{4 \pi} \right)^{1/3}$ and $(r+h) = \left( \frac{3 \, (\volcav + \volmyo)}{4 \pi} \right)^{1/3}$.
Using the volume-based representations of $r$ and $r+h$, we can rewrite Eq.~\eqref{supp:equ:app:lol:ext}
\begin{align}
  \sigmaLext &= \frac{p \, r}{2 \, h \left( 1 + \frac{h}{2 \, r} \right) } = \frac{p}{\frac{2 \, h}{r} \left( 1 + \frac{h}{2 \, r} \right) }
  = \frac{p}{\left( \frac{2 \, h}{r} + \frac{2 \, h^2}{2 \, r^2} \right) }
  = \frac{p}{\left( \frac{r^2}{r^2} + \frac{2 \, h \, r}{r^2} + \frac{h^2}{r^2} \right) - 1}
  = \frac{p}{\left( \frac{r+h}{r} \right)^2 - 1} \nonumber \\
  &= \frac{p}{\left( \frac{\volcav + \volmyo}{\volcav} \right)^{2/3} - 1} = \sigmaLvol
\end{align}
which we denote as $\sigmaLvol$.


\subsection{Computation of power and work}
\label{supp:subsec:power}

The approximations $\sigmaLsmp$, $\sigmaLext$ and $\sigmaLvol$ for the circumferential stress $\sigmacirc$
can be used to derive an estimator for the internal power
\begin{equation}
  \label{supp:equ:ipower}
  \powerint(t) = \int \limits_{\Omega} \sigmatot(\vu, t) : \straindot (\vu, t) \dvx
\end{equation}
and internal work
\begin{equation}
  \label{supp:equ:iwork}
  \workint = \int \limits_{t_0}^{t} \powerint(\tau) d\tau.
\end{equation}
For this sake, we consider Eq.~\eqref{supp:equ:ipower} and the simplified representation of
the total stress tensor Eq.~\ref{supp:equ:app:sig:sph:smp}. In Eq.~\eqref{supp:equ:ipower}
an approximation of the strain rate $\straindot$ is required. Rewriting the strain tensor
$\strain$ in spherical coordinates, as done for the stress tensor $\sigmatot$, we obtain
\begin{displaymath}
  \strain = \left(
  \begin{array}{ccc}
  \varepsilon_{xx} & \varepsilon_{xy} & \varepsilon_{xz} \\
  \varepsilon_{yx} & \varepsilon_{yy} & \varepsilon_{yz} \\
  \varepsilon_{zx} & \varepsilon_{zy} & \varepsilon_{zz}
  \end{array}
  \right) = \mathbf{P} \left(
  \begin{array}{ccc}
  \varepsilon_{r r} & \varepsilon_{r \varphi} & \varepsilon_{r \theta} \\
  \varepsilon_{\varphi r} & \varepsilon_{\varphi \varphi} & \varepsilon_{\varphi \theta} \\
  \varepsilon_{\theta r} & \varepsilon_{\theta \varphi} & \varepsilon_{\theta \theta} \\
  \end{array}
  \right) \mathbf{P}^{\top}
\end{displaymath}
where $\mathbf{P}$ is the projection matrix introduced in Sec.~\ref{supp:subsec:stresses}.
Similarly, the strain rate $\straindot$ is expressed as
\begin{displaymath}
\straindot = \mathbf{P} \left(
\begin{array}{ccc}
\dot{\varepsilon}_{r r} & \dot{\varepsilon}_{r \varphi} & \dot{\varepsilon}_{r \theta} \\
\dot{\varepsilon}_{\varphi r} & \dot{\varepsilon}_{\varphi \varphi} & \dot{\varepsilon}_{\varphi \theta} \\
\dot{\varepsilon}_{\theta r} & \dot{\varepsilon}_{\theta \varphi} & \dot{\varepsilon}_{\theta \theta} \\
\end{array}
\right) \mathbf{P}^{\top}.
\end{displaymath}
Using Eq.~\eqref{supp:equ:app:sig:sph:smp}, an approximation of the internal power density, $\powerintdensity$
can be derived as
$(\sigmatot : \straindot) \approx \sigmacirc \left( \dot{\varepsilon}_{\varphi \varphi} + \dot{\varepsilon}_{\theta \theta}\right)$.
Due to the assumption of symmetry \textbf{(A2)}, strains in circumferential
direction do not vary with space, i,e.
$\varepsilon_{\varphi \varphi} = \varepsilon_{\theta \theta} = \txtsubscript{\varepsilon}{circ}$,
and the approximation for the internal power density simplifies to
\begin{equation}
  (\sigmatot : \straindot) \approx 2 \, \sigmacirc \, \txtsubscript{\dot{\varepsilon}}{circ}.  \label{supp:equ:app:powerdensity}
\end{equation}

\begin{figure}[!ht]
  \centering
  \subfloat[\;Circumferential Strain.]{ \includegraphics[scale=0.7]{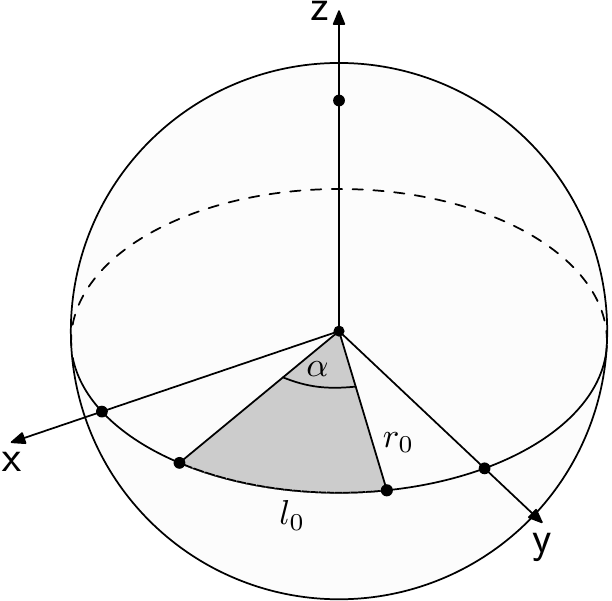} }
  \hspace{1.5cm}
  \subfloat[\;Initial (left) and deformed (right) configuration.]{ \includegraphics[scale=0.7]{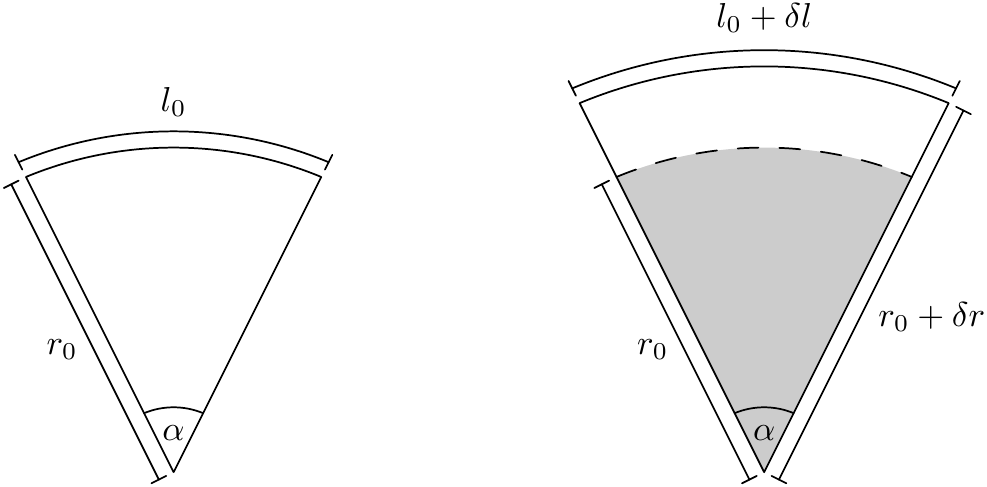} }
  \caption{Strain estimation.}
  \label{supp:fig:strain}
\end{figure}

An approximation of the circumferential strain $\txtsubscript{\varepsilon}{circ}$
can be found based the Cauchy strain and considerations illustrated in Fig.~\ref{supp:fig:strain}.
Accordingly, for a given radius $r$ circumferential strain can be approximated as
\begin{displaymath}
  \txtsubscript{\varepsilon}{circ} \approx \frac{(l_0+\delta l) - l_0}{l_0} =
  \frac{(r_0+\delta r) \, \alpha - r_0 \, \alpha}{r_0 \, \alpha} =
  \frac{(r_0+\delta r) - r_0}{r_0} = \frac{r}{r_0} - 1
\end{displaymath}
and for the circumferential strain rate we obtain $\txtsubscript{\dot{\varepsilon}}{circ} \approx \frac{\dot{r}}{r_0}$.

To approximate the circumferential strain rate $\txtsubscript{\dot{\varepsilon}}{circ}(\vu, t)$ of a spherical shell of
thickness $h = R - r$, we take the arithmetic mean of the strain rate at inner radius $r$ and outer radius $R$, that is
\begin{equation}
  \txtsubscript{\dot{\varepsilon}}{circ}(\vu, t) \approx \frac{1}{2} \left( \frac{\dot{r}(t)}{r_0} + \frac{\dot{R}(t)}{R_0} \right) \label{supp:equ:app:strainrate}
\end{equation}
where $r_0$ is the initial inner radius and $R_0$ is the initial outer radius
of the spherical shell at its stress free configuration, i.e. $p = 0$.

Using Eqs.~\eqref{supp:equ:app:powerdensity} and \eqref{supp:equ:app:strainrate}, the internal power can be estimated by
\begin{displaymath}
  \begin{split}
    \powerint(t) & = \int \limits_{\Omega} \sigmatot(\vu, t) : \straindot(\vu, t) \dvx
    \approx \volmyo(t) \, 2 \, \sigmacirc(\vu, t) \, \frac{1}{2} \left( \frac{\dot{r}(t)}{r_0} + \frac{\dot{R}(t)}{R_0} \right) \\
    & = \volmyo(t) \, \sigmacirc(\vu, t) \, \left( \frac{\dot{r}(t)}{r_0} + \frac{\dot{R}(t)}{R_0} \right)
  \end{split}
\end{displaymath}
where $\volmyo(t)$ is the volume of the shell's wall at time $t$.

Using an estimate for $\sigmacirc$ we obtain
\begin{displaymath}
  \powerintstar(t) = \volmyo(t) \, \sigmaLstar(t) \, \left( \frac{\dot{r}(t)}{r_0} + \frac{\dot{R}(t)}{R_0} \right)
\end{displaymath}
with $\star \in \left\{ \text{h}, \text{H}, \text{V} \right\}$ and by integrating over time, we get an estimate
for the internal work
\begin{displaymath}
  \workintstar = \int \limits_{0}^{T} \powerintstar(t) \dt.
\end{displaymath}

\subsection{Methods}

\subsubsection{Model fitting}

To delineate anisotropy from pure geometry effects,
passive inflation experiments were also performed with LV models using the Demiray model
and passive mechanical behavior was compared to the spherical shell models \sphh, \sphH and \sphHH.
In these cases, parameters were set to $b = 7$ and $a$ was chosen in a patient-specific manner
to obtain the same volume at maximum inflation pressure
as with the Guccione model. 
Note that in none of the simulations of a full cardiac cycle
the Demiray model was considered as the resulting kinematics was in stark contrast to the clinical data.

\subsubsection{Analysis of LV inflation experiments}

To evaluate the influence of violating the assumption on the geometry \textbf{(A2)},
passive inflation experiments were performed with LV models and the isotropic material
due to Demiray, Eq.~\eqref{eq:demirayStrainEnergy}, following the same protocol as applied
to the spherical shell models $\sphh$, $\sphH$ and $\sphHH$. 
The parameter $a$ in the Demiray model was set to $0.45$, $0.63$, $0.41$ and $0.38$ kPa
for the models $\lva$, $\lvb$, $\lvc$ and $\lvd$, respectively. The Laplace-based stress
estimates $\sigmaLsmp$, $\sigmaLext$ and $\sigmaLvol$ were compared to the mean stresses
obtained from the FE solution. Stresses were evaluated with respect to an ellipsoidal coordinate system
to facilitate a comparison with stresses computed in the spherical shell models
$\sphh$, $\sphH$ and $\sphHH$ where spherical coordinates were used for stress analysis.
The ellipsoidal coordinate system for the LV models was constructed
by assigning fiber and sheet orientations using a rule-based method
with a constant fiber angle of $0^\circ$. Stress components $\sigmarad(\vx)$,
$\sigmaphi(\vx)$ and $\sigmatht(\vx)$ were averaged yielding $\msigmarad$, $\msigmaphi$
and $\msigmatht$, respectively. Note that all models except $\sphh$ showed marked spatial
stress variations. Thus, the reported mean stresses $\bar{\sigma}$ may deviate considerably
from the true local stresses $\sigma(\vx)$. Laplace-based estimations of power $\powerintstar$ and work $\workintstar$,
were compared to those obtained by FE simulation, $\powerint$ and $\workint$
and to external hydrodynamic power and work in the LV cavity, $\powerext$ and $\workext$.

\subsection{Results}

\subsubsection{Verification of the FE model}

Similarly, with increasing $h$ the accuracy of the thick-walled Laplace estimate $\workintext$
performed better than the simpler thin-walled Laplace estimate $\workintsmp$.
As expected on grounds of conservation of energy, the agreement between biomechanical work
$\workint$ and hemodynamic work $\workext$ was essentially perfect with differences $<2\%$ for all models.

\subsubsection{Passive inflation of LV models}

\begin{figure}[!h]
  \centering
  \includegraphics[width=1.0\textwidth, keepaspectratio]{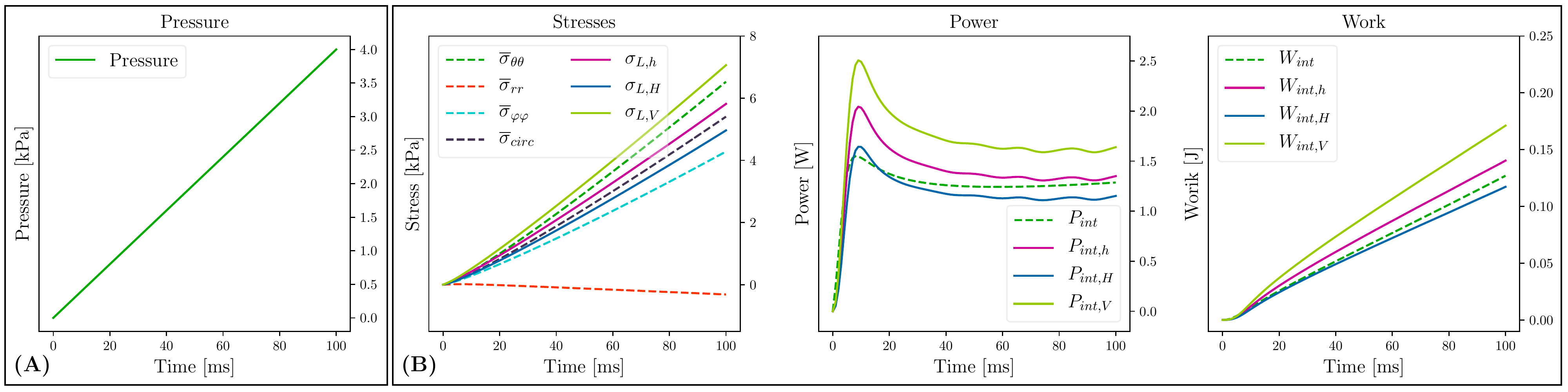}
  \caption{(A) Loading protocol. (B) Stresses, power and work for isotropic model.
    Data are shown for model $\lvd$.}
  \label{supp:fig:lv_to_laplace}
\end{figure}

The LV models $\lva$--$\lvd$ were inflated following the loading protocol in Fig.~\ref{supp:fig:lv_to_laplace}.A.
Passive material behavior was represented compliant with \textbf{(A1)} by the isotropic Demiray model.
The temporal evolution of FE- and Laplace-based stresses, power and work
are shown in Fig.~\ref{supp:fig:lv_to_laplace}.B for model $\lvd$.
Minor quantitative differences to other models $\lva$--$\lvc$ were observed,
but qualitatively the overall behavior was identical.
Stresses at $p=4$~kPa and the amount of work incurred during inflation up to this pressure
are summarized in Tabs.~\ref{supp:tab:exp:stress} and ~\ref{supp:tab:exp:work}.

\begin{center}
  \begin{table}[h!]
    \centering
    \resizebox{\linewidth}{!}{
    \begin{tabular*}{500pt}{@{\extracolsep\fill}l|ccc|ccc|cc@{\extracolsep\fill}}
      \toprule
      Setup & $\msigmarad$ [kPa] & $\msigmatht$ [kPa] & $\msigmaphi$ [kPa] & $\sigmaLsmp$ [kPa] & $\sigmaLext$ [kPa] & $\sigmaLvol$ [kPa]& \#elements & $\bar{\dx}$ [mm] \\
      \midrule
      $\lvadem$ & $-0.39$ & $6.56$ & $3.77$ & $5.16$ & $4.32$ & $6.81$ & $420704$ & $1.52$ \\
      $\lvbdem$ & $-0.38$ & $8.26$ & $4.78$ & $5.92$ & $5.07$ & $8.24$ & $332221$ & $1.74$ \\
      $\lvcdem$ & $-0.21$ & $4.07$ & $2.26$ & $3.43$ & $2.65$ & $4.41$ & $456553$ & $1.84$ \\
      $\lvddem$ & $-0.31$ & $6.53$ & $4.28$ & $5.81$ & $4.69$ & $7.06$ & $394808$ & $1.86$ \\
      \bottomrule
  \end{tabular*}}
    \caption{Comparison of FE-based mean wall stresses $\msigmarad$, $\msigmatht$ and $\msigmaphi$
      in radial, azimuthal and meridional direction, respectively,
      with the Laplace-based wall stress estimates $\sigmaLsmp$, $\sigmaLext$
      and $\sigmaLvol$. All stresses refer to the maximum applied pressure of $p=4~$kPa.}
    \label{supp:tab:exp:stress}
  \end{table}
\end{center}
\vspace{-0.8cm}


\begin{center}
  \begin{table}[h!]
    \centering
    \resizebox{\linewidth}{!}{
    \begin{tabular*}{500pt}{@{\extracolsep\fill}l|cc|ccc|cc@{\extracolsep\fill}}
      \toprule
      Setup & $\workext$ [mJ] & $\workint$ [mJ] & $\workintsmp$ [mJ] & $\workintext$ [mJ] & $\workintvol$ [mJ] & \#elements & $\bar{\dx}$ [mm] \\
      \midrule
      $\sphh$ & $16.82$ & $16.92$ & $19.86$ & $19.65$ && $83825$  & $0.65$ \\
      $\sphH$ & $10.94$ & $10.99$ & $11.58$ & $10.87$ && $40974$  & $1.23$ \\
      $\sphHH$ & $8.30$ & $8.15$ & $10.89$ & $7.51$ && $54449$  & $2.31$ \\
      \midrule
      $\lvadem$ & $67.03$  & $65.48$  & $67.94$  & $55.67$  & $89.30$  & $420704$ & $1.52$ \\
      $\lvbdem$ & $111.67$ & $110.01$ & $124.41$ & $104.31$ & $183.99$ & $332221$ & $1.74$ \\
      $\lvcdem$ & $96.19$  & $94.90$  & $100.48$ & $75.44$  & $128.44$ & $456553$ & $1.84$ \\
      $\lvddem$ & $129.19$ & $126.98$ & $140.30$ & $117.23$ & $171.12$ & $394808$ & $1.86$ \\
      \midrule
      $\lvagu$ & $93.19$  & $89.34$  & $89.11$  & $72.65$  & $116.26$ & $420704$ & $1.52$ \\
      $\lvbgu$ & $149.55$ & $145.27$ & $167.23$ & $131.08$ & $219.04$ & $332221$ & $1.74$ \\
      $\lvcgu$ & $136.84$ & $133.42$ & $146.17$ & $109.42$ & $183.75$ & $456553$ & $1.84$ \\
      $\lvdgu$ & $182.77$ & $177.51$ & $173.92$ & $143.91$ & $216.12$ & $394808$ & $1.86$ \\
      \bottomrule
  \end{tabular*}}
    \caption{
      Comparison of FE-based biomechanical and hemodynamic work,
      $\workint$ and $\workext$,
      with the Laplace-based work estimates $\workintsmp$, $\workintext$ and $\workintvol$
      for passive inflation with a pressure of $p=4$ kPa.}
    \label{supp:tab:exp:work}
  \end{table}
\end{center}

\subsubsection{Analysis of LV cycle experiments}

\begin{figure}[!h]
  \centering
  \includegraphics[width=1.0\textwidth, keepaspectratio]{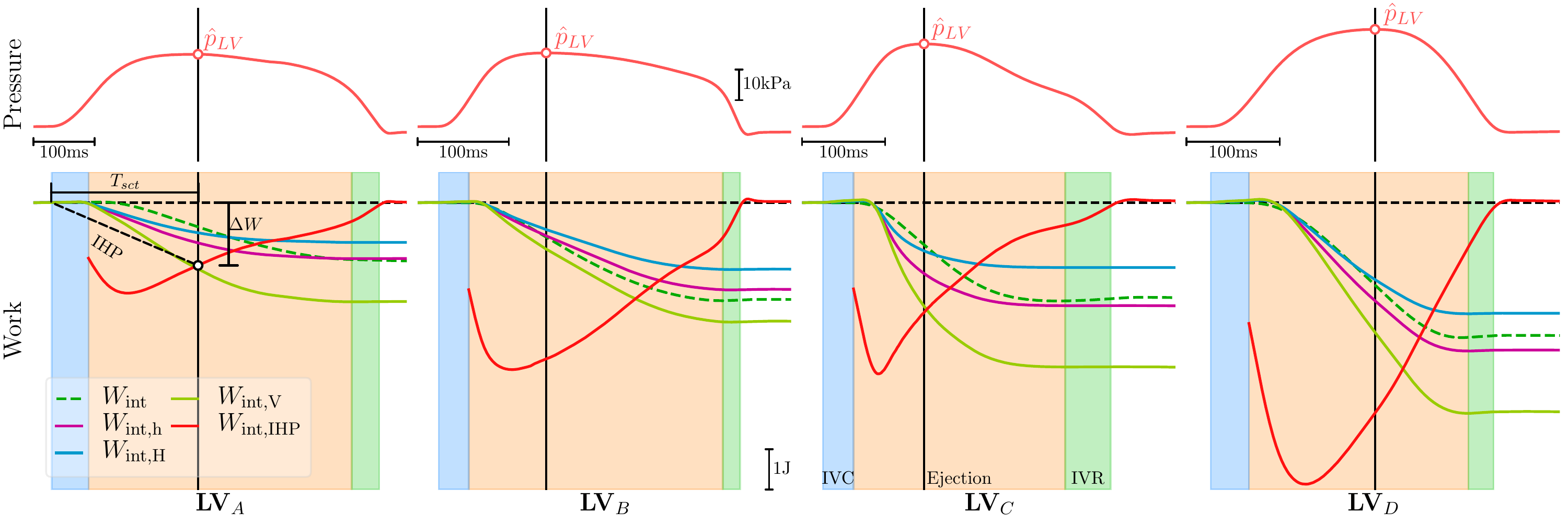}
  \caption{Comparison of FE-based computation $\workint$
    with Laplace-based estimates of work $\workintsmp$,  $\workintext$ and  $\workintvol$.
    Top panels show the time course of pressure $p$ in the LV endocardium.
    The solid black vertical line indicates the instant, $\tplvpeak$, when peak pressure in the LV, $\plvpeak$, occurs.}
  \label{supp:fig:_working_lv_power:work_only}
\end{figure}

\begin{figure}[!h]
  \centering
  \includegraphics[scale=0.75]{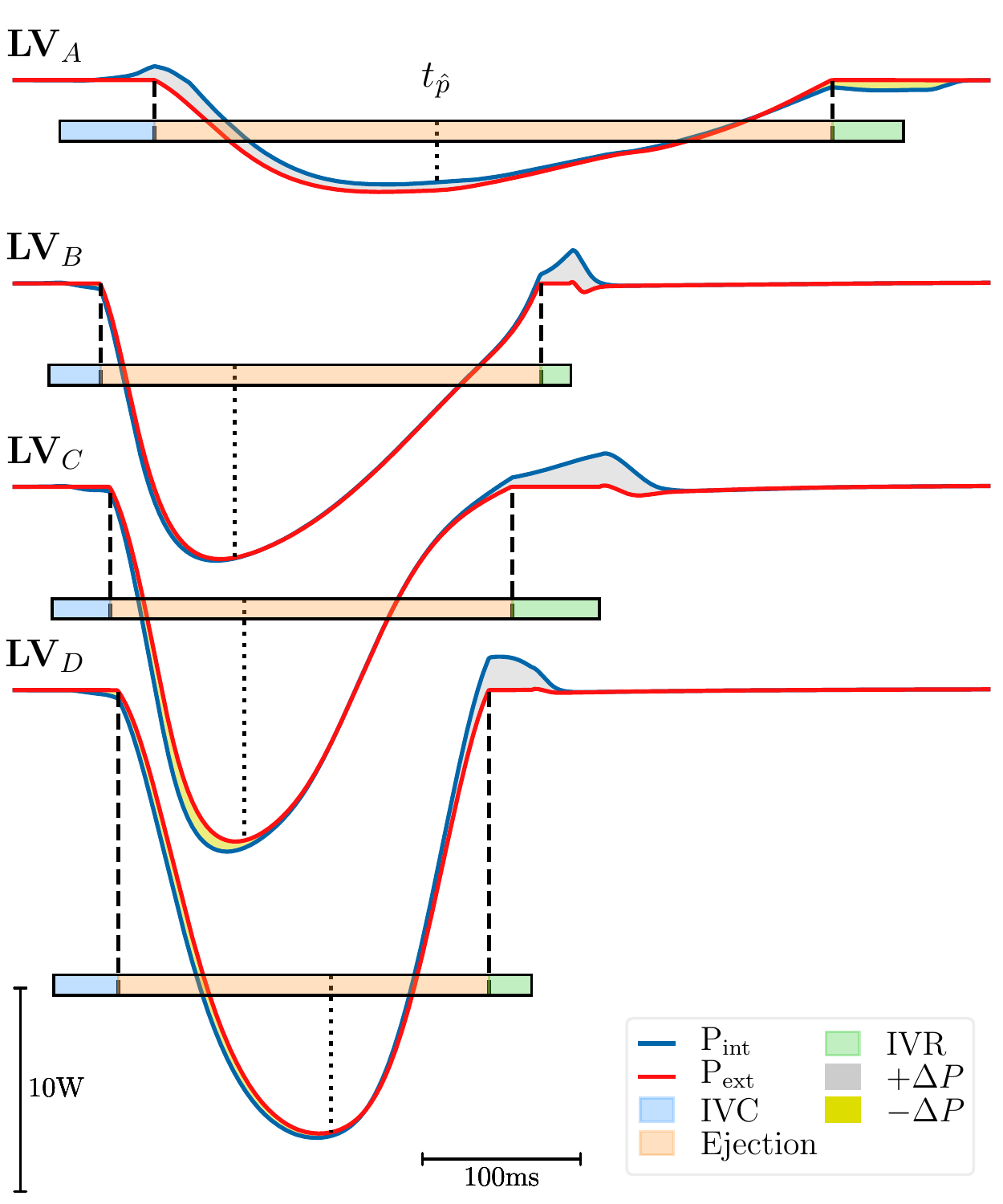}
  \caption{Differences between mechanical versus hydrodynamic power,
    $\Delta P = \powerint - \powerext$, during IVC and early ejection
    were very minor (dark yellow area). A slightly more pronounced $\Delta P$
    is witnessed during late ejection and IVR (gray area).}
  \label{supp:fig:int_vs_ext_power}
\end{figure}

\end{document}